\documentclass[times,final]{elsarticle}
\usepackage[margin=2.5cm]{geometry}

\usepackage{framed,multirow}

\usepackage{amssymb}
\usepackage{latexsym}

\usepackage{url}
\usepackage{xcolor}
\definecolor{newcolor}{rgb}{.8,.349,.1}

\usepackage[utf8]{inputenc}
\usepackage{url}
\usepackage{array}
\usepackage{amsmath,amssymb,amsfonts}
\usepackage{mathtools}
\usepackage{algorithm}
\usepackage{algpseudocode}
\usepackage{tikz}
\usepackage{graphicx}
\usepackage{subcaption}
\usepackage[export]{adjustbox}
\usepackage{bm}
\usepackage{xcolor}

\begin{document}


\begin{frontmatter}

\title{Bridging the computational gap between mesoscopic and continuum modeling of red blood cells for fully resolved blood flow}

\author[cui]{Christos Kotsalos \corref{cor1}}
\ead{christos.kotsalos@unige.ch}
\cortext[cor1]{Corresponding author: Tel.: +41-22-379-0169}

\author[cui]{Jonas Latt}
\ead{jonas.latt@unige.ch}

\author[cui]{Bastien Chopard}
\ead{bastien.chopard@unige.ch}

\address[cui]{Computer Science Department, University of Geneva, 7 route de Drize, CH-1227 Carouge, Switzerland}


\begin{abstract}
We present a computational framework for the simulation of blood flow with fully resolved red blood cells (RBCs) using a modular approach that consists of a lattice Boltzmann solver for the blood plasma, a novel finite element based solver for the deformable bodies and an immersed boundary method for the fluid-solid interaction. For the RBCs, we propose a nodal projective FEM (npFEM) solver which has theoretical advantages over the more commonly used mass-spring systems (mesoscopic modeling), such as an unconditional stability, versatile material expressivity, and one set of parameters to fully describe the behavior of the body at any mesh resolution. At the same time, the method is substantially faster than other FEM solvers proposed in this field, and has an efficiency that is comparable to the one of mesoscopic models. At its core, the solver uses specially defined potential energies, and builds upon them a fast iterative procedure based on quasi-Newton techniques. For a known material, our solver has only one free parameter that demands tuning, related to the body viscoelasticity. In contrast, state-of-the-art solvers for deformable bodies have more free parameters, and the calibration of the models demands special assumptions regarding the mesh topology, which restrict their generality and mesh independence. We propose as well a modification to the potential energy proposed by Skalak et al. 1973 for the red blood cell membrane, which enhances the strain hardening behavior at higher deformations. Our viscoelastic model for the red blood cell, while simple enough and applicable to any kind of solver as a post-convergence step, can capture accurately the characteristic recovery time and tank-treading frequencies. The framework is validated using experimental data, and it proves to be scalable for multiple deformable bodies.
\end{abstract}


\end{frontmatter}


\section{Introduction}
Blood is a complex suspension of red blood cells (RBCs), white blood cells and platelets, submerged in a Newtonian fluid, the plasma. Especially, the RBCs (else, erythrocytes) cover 40-45\% of the whole blood volume, and their behavior has a direct impact on blood rheology. The accurate modeling of the collective transport of the cells in the plasma is of paramount importance since it can help us decipher \emph{in vivo} phenomena, e.g., blood clotting, margination of platelets, or the cell-free layer (Mountrakis 2015 \citep{Mountrakis2015TransportModels}). In addition, the deformability of RBCs is strongly linked to some pathological conditions, e.g., hereditary disorders (like spherocytosis, elliptocytosis, and stomatocytosis), metabolic disorders (like diabetes, hypercholesterolemia, and obesity), malaria, or sickle anemia as described by Tomaiuolo 2014 \citep{Tomaiuolo2014BiomechanicalMicrofluidics}. In more details, a human RBC has a biconcave discocyte shape with a liquid interior (cytoplasm, hemoglobin solution) of dynamic viscosity $6-10~cP$. A healthy red blood cell has an average surface area of $134~\mu m^2$, a volume of $94~\mu m^3$, a diameter of $7.82~\mu m$ and a varying thickness of $0.81~\mu m$ at the dimple to $2.57~\mu m$ at the periphery. An analytical formula for the RBC shape is proposed in Evans and Skalak 1980 \citep{Evans1980MechanicsBiomembranes}. The RBC membrane is a complex multi-layered structure consisting of an external lipid bilayer attached to an underlying cytoskeleton (spectrin-link network). These complicated anucleate structures transfer oxygen between blood and tissues thanks to their ability to undergo substantial deformations while maintaining their area and volume.

The absence of a universal numerical model capable of describing the viscoelastic behavior of a RBC is the motivation of this study. A universal model should fulfill a number of criteria, such as generality, robustness, accuracy and performance. By generality and accuracy, we mean the ability of the model to simulate any material and all the expected behavior as close as possible to the observed data (experiments). By robustness, we refer to the capability of the model to adapt to extreme physical circumstances such as large deformations, confined flows, or high flow shear rates, and to numerically unfavorable settings like large time steps. Last but not least, performance is a key requirement in the field of computational biophysics, since the model should be extended to cases including millions or even billions of deformable bodies. To give an example, a volume of $1~mm^3$ of blood at a hematocrit 40\% contains approximately five million RBCs. For the modeling of RBCs, two main approaches are proposed in the literature. The first is a continuum-level approach, that views the membrane as a continuous medium which obeys specific partial differential equations and constitutive laws. The discretization of these equations is commonly done through a finite element method (FEM). Many researchers have successfully used this approach for the modeling of membranes, including Kr{\"{u}}ger et al. 2011 \citep{Kruger2011EfficientMethod}, MacMeccan et al. 2009 \citep{Macmeccan2009SimulatingMethod}, Shrivastava and Tang 1993 \citep{Shrivastava1993LargeThermoforming}. Kl\"oppel et al. 2011 \citep{Kloppel2011AErythrocytes} employed a two-layer model for an even more accurate description of the lipid bilayer and the underlying cytoskeleton. The aforementioned models use constitutive laws for the description of the membrane material, such as the material introduced by Skalak et al. 1973 \citep{Skalak1973StrainMembranes}, Neo-Hookean materials proposed in \citep{Bonet2008NonlinearAnalysis} or the material described by Yeoh 1990 \citep{Yeoh1990CharacterizationVulcanizates} (for a thorough investigation of material models the reader is referred to Dimitrakopoulos et al. 2012 \citep{Dimitrakopoulos2012AnalysisModeling} and Siguenza et al. 2017 \citep{Siguenza2017HowMechanics}). {\color{black} Among the many advantages of this approach is the guaranteed mesh-independence, which is a product of the way the equations are discretized}. This modeling approach satisfies many of the above-mentioned criteria, namely generality, robustness and accuracy. However, they do not perform well in simulations of multiple deformable bodies due to their high computational cost. The second modeling approach operates at a mesoscopic level, and represents surface properties through a mass-spring system. More precisely, mesoscopic modeling of the RBCs spans from the spectrin-link to the coarse-grained spectrin-link levels. The spectrin-level approach tries to imitate the physics of the cytoskeleton of a RBC as a fine-grained mass-spring model. The spectrin level was introduced by Discher et al. 1998 \citep{Discher1998SimulationsAspiration} and extended further by Dao et al. 2003 \& 2006 \citep{Dao2003MechanicsTweezers, Dao2006MolecularlyErythrocyte} and Li et al. 2005 \citep{Li2005Spectrin-levelErythrocyte}. Though, this approach is limited by the high computational cost, given that it requires $23~867$ surface vertices \citep{Pivkin2008AccurateCells} for the simulation of the whole spectrin network. For this reason, the coarse-grained models of Dupin et al. 2007 \citep{Dupin2007ModelingDimensions}, Pivkin and Karniadakis 2008 \citep{Pivkin2008AccurateCells}, Fedosov et al. 2010 \citep{Fedosov2010ARheologydynamics}, and Reasor et al. 2011 \citep{Reasor2011CouplingFlow} have gained substantial attention during the last years, where the surface discretization demands about $500$ vertices. Currently, state-of-the-art solvers like Hemocell \citep{Zavodszky2017CellularCells, Zavodszky2017Hemocell:Library} and the solver of Fedosov \citep{Fedosov2010ARheologydynamics} are based on this modeling approach. These coarse-grained methods outperform all the others in terms of computational time. Nevertheless, since they are mass-spring systems they inherit limitations on generality, robustness and accuracy.

As far as the simulation of the blood plasma is concerned, there exists a plethora of mature CFD approaches. For our study, we make use of the lattice Boltzmann method (LBM) \citep{Kruger2017TheMethod} which indirectly solves the Navier-Stokes equations. Another candidate method widely used for the simulation of blood flows, is dissipative particle dynamics (DPD) \citep{Pivkin2008AccurateCells, Fedosov2010ARheologydynamics}. Both are particle-based methods which deploy either collision and streaming operations or Newtonian laws, respectively, for the time advancement of the fluid.

The coupling of the fluid with the solid phase is a critical point for an accurate and stable simulation. Peskin 1972 \citep{Peskin1972FlowMethod} developed the immersed boundary method (IBM) to model blood flow in combination with moving heart valves. The strength of the IBM is that the fluid solver does not have to be modified except from the addition of a forcing term, and the fluid mesh does not need to be adjusted dynamically. Moreover, the deformable body and its discrete representation do not need to conform to the discrete fluid mesh, which leads to a very efficient fluid-solid coupling. In the context of lattice Boltzmann, an alternative solution would be the deployment of moving bounce back nodes \citep{Kruger2017TheMethod}. With this approach, the computational cost however steeply increases, given the need to track the lattice nodes that transition from solid to fluid state.

In a typical numerical framework for blood flow, the computational time is dominated by the structural solver for the deformable RBCs. Even with the relatively cheap mass-spring systems,  Z\'avodsky et al. 2017 \citep{Zavodszky2017Hemocell:Library} report for example that the deformable bodies solver constitutes over 95\% of the total computational time. Since the numerical models for Newtonian fluid flow and for the fluid-solid coupling are more mature than the ones for the physics of RBCs, our main focus is to develop a novel solution which strictly fulfills the criteria of generality, robustness, accuracy and performance without any compromise. We propose a novel approach for deformable viscoelastic bodies based on nodal projective finite elements method (npFEM). Inspired by the research of Liu et al. 2017 \citep{Liu2017Quasi-NewtonMaterials} and Bouaziz et al. 2014 \citep{Bouaziz2014ProjectiveSimulation}, our solver extends these theories to the field of computational biophysics. The expression ``nodal'' refers to the mass lumping technique, in which both the masses and the forces are lumped on the vertices of the discretized body, and therefore the finite elements act like generalized viscoelastic springs. The term ``projective'' stands for the use of specially designed potential energies that help us build a fast solver based on quasi-Newton optimization techniques \citep{Bathe1980SomeEquations, Fish1995AnSystems}. Lastly, we discretize the continuous potential energies using linear basis functions, leading to the standard Galerkin method \citep{Quarteroni2009NumericalProblems} in the field of FEM. Our solver inherits the versatility and robustness of FEM and is almost as fast as plain mass-spring systems. To our knowledge, this is the first solver bridging the gap between the two approaches, continuum and mesoscopic, in the field of RBC modeling.

{\color{black}
The present work is organized as follows: In section \ref{sec:METHODS} we thoroughly introduce the npFEM method focusing on the modeling of RBCs, the lattice Boltzmann method and the immersed boundary method. Following, in section \ref{sec:RECAP} we put all the different pieces together and show how the complete computational framework operates. Finally, in section \ref{sec:Results} we present extensive validations of our computational framework and give insights about its performance and scalability to systems of multiple blood cells.
}

\section{Methods} \label{sec:METHODS}
The focus of this study is to build a computational framework for fully resolved 3D blood flow. The constituents of this framework are the solid body solver ({\color{black}section \ref{sec:npFEM}}), the fluid solver ({\color{black}section \ref{sec:LBM}}), and the fluid-structure interaction (FSI) ({\color{black}section \ref{sec:IBM}}). The solid body solver based on the nodal projective finite elements method (npFEM) is our main contribution to the field. Given the volume fraction of red blood cells compared to the other cells (white, platelets), and consequently their importance on blood rheology, we develop the npFEM solver around RBC simulations. The fluid solver, which is used for the simulation of the blood plasma, is based on the lattice Boltzmann method (LBM) as implemented in the open source software Palabos \citep{PalabosHttp://www.palabos.org}. The coupling between the two solvers is done via an immersed boundary method known as multidirect forcing scheme \citep{Ota2012LiftSimulations}.

\subsection{Nodal projective FEM (npFEM)} \label{sec:npFEM}
This section provides a complete review of the methods introduced in the research of Liu et al. 2017 \citep{Liu2017Quasi-NewtonMaterials} and Bouaziz et al. 2014 \citep{Bouaziz2014ProjectiveSimulation}, and some novel extensions proposed to adapt them to the field of RBC modeling. A validation of these adaptations is provided in later sections, see results section \ref{sec:Results}. A red blood cell is represented by its triangulated membrane, and its total mass is lumped on the membrane vertices (including the interior fluid). This section presents how we realize the dynamics of deformable bodies, as well as all the potential energies that describe the behavior of a red blood cell, namely local area and global volume conservation, bending and material.

Let us assume a surface mesh consisting of $n$ vertices with positions $\bm{x} \in \mathbb{R}^{n \times 3}$ and velocities $\bm{v} \in \mathbb{R}^{n \times 3}$. The evolution of the system in time follows Newton's law of motion. The external forces are defined as $\bm{F}_{ext} \in \mathbb{R}^{n \times 3}$ and come from the fluid-solid interaction, while the internal forces are $\bm{F}_{int} \in \mathbb{R}^{n \times 3}$. The internal forces are position dependent, i.e., $\bm{F}_{int}(\bm{x}) = - \sum_{i} \nabla E_i(\bm{x})$, where $E_i(\bm{x})$ is a scalar discrete elemental potential energy. Moreover, the summation of all the elemental potential energies of any type results in the total elastic potential energy of the body. Conservative internal forces, i.e, derived from potentials, are closely related to the assumption of hyperelastic materials \citep{Sifakis:2012:FSD:2343483.2343501}. The \emph{implicit Euler time integration} leads to the following advancement rules (subscripts $n$ and $n+1$ refer to time $t$ and $t+h$, respectively):
\begin{align}
    \bm{v}_{n+1} = \frac{\bm{x}_{n+1} - \bm{x}_{n}}{h},    \\
    \bm{F}_{int}(\bm{x}_{n+1}) + \bm{F}_{ext}(\bm{x}_{n}) - \mathbf{C} \bm{v}_{n+1} = \mathbf{M} \frac{\bm{v}_{n+1} - \bm{v}_{n}}{h}, \label{eq:RayleighNewton}
\end{align}
where $\mathbf{M} \in \mathbb{R}^{n \times n}$ is the mass matrix, $h$ is the time step and $\mathbf{C} \in \mathbb{R}^{n \times n}$ is the damping matrix acting like a proxy for viscoelasticity (Rayleigh damping, see section \ref{sec:Viscoelasticity}). The mass matrix is built by lumping the total mass of the body on the mesh vertices, resulting in a diagonal structure. The lumping can be done either by equally distributing the mass or by weighting, based on the corresponding area per vertex. Combining the above equations we derive
\begin{equation} \label{eq:preOptimizationNewton}
    \widetilde{\mathbf{M}} (\bm{x}_{n+1} - \bm{y}_n) = h^2 \bm{F}_{int}(\bm{x}_{n+1}),
\end{equation}
where $\widetilde{\mathbf{M}} = \mathbf{M} + h \mathbf{C}$ and $\bm{y}_n = \bm{x}_{n} + h \widetilde{\mathbf{M}}^{-1} \mathbf{M} \bm{v}_n + h^2 \widetilde{\mathbf{M}}^{-1} \bm{F}_{ext}$. Equation \eqref{eq:preOptimizationNewton} can be turned into an optimization problem \citep{Bouaziz2014ProjectiveSimulation} as
\begin{equation} \label{eq:genericOptimizationNewton}
    \underset{\bm{x}_{n+1}}{\min}~~\frac{1}{2h^2} \left \| \widetilde{\mathbf{M}}^{\frac{1}{2}} (\bm{x}_{n+1} - \bm{y}_n) \right \| ^{2}_{F} + \sum_{i} E_i(\bm{x}_{n+1}),
\end{equation}
where $|| \cdot ||_F$ is the Frobenius norm. Indeed, setting the derivative of equation \eqref{eq:genericOptimizationNewton} to zero (thereby minimizing the objective function), we recover Newton's second law of motion. The solution to the minimization problem gives $\bm{x}_{n+1}$. The choice of implicit Euler integration serves a dual purpose. Firstly, it provides unconditional stability for arbitrary time step. Secondly, this integration scheme is characterized by some numerical dissipation linked to the time step (the smaller the time step the smaller the numerical dissipation). This dissipation enhances the stability of the coupling (fluid \& solid) by reducing energy oscillations that could potentially lead to instabilities. This dissipation term acts like the viscous dissipation provided by Rayleigh damping (matrix $\mathbf{C}$).

{\color{black}
A remark at this point is that the variational form of implicit Euler \eqref{eq:genericOptimizationNewton} is just a reformulation of Newtonian dynamics, which in their turn can be viewed from the Hamiltonian dynamics perspective. Nevertheless, to accurately capture the trajectory and deformation of the bodies (dynamics) we need to guarantee the conservation of linear \& angular momenta. For this, one has to be cautious with the selection of the potential energies and the viscoelastic terms. According to Noether's theorem, momenta are conserved when the elastic energies are rigid motion invariant. For the viscoelastic terms, we give an extended discussion on the conservation of momenta in section \ref{sec:Viscoelasticity}. Taking into consideration all the above, we guarantee that we resolve as accurately as possible the dynamics of the simulated bodies, reproducing the correct physics as dictated by Newton's laws.
}

\subsubsection{Extended projective dynamics method and quasi-Newton optimization}
Projective dynamics method \citep{Bouaziz2014ProjectiveSimulation} requires a special form of potential energies, based on the notion of constraint projection:
\begin{equation} \label{eq:PDEnergy}
    E_i(\bm{x}) = \underset{\mathbf{S}_i \bm{p} \in \mathcal{M}_i}{\min} \widetilde{E}_i(\bm{x}, \mathbf{S}_i \bm{p}),~~\widetilde{E}_i(\bm{x}, \bm{z}) = \left \| \mathbf{G}_i \bm{x} - \bm{z} \right \|^{2}_{F},
\end{equation}
where $\mathcal{M}_i$ is a constraint manifold defined by a desired undeformed state, $\bm{p}_i = \mathbf{S}_i \bm{p} \in \mathbb{R}^{m \times 3}$ is an auxiliary projection variable on $\mathcal{M}_i$ and $\mathbf{G}_i$ is a mapping from $\bm{x}$ to an element-wise deformation representation which depends on the mesh topology and the initial configuration. For an elastic body, there are many elemental energies of various types, therefore various projections. For compactness, we stack all the projections into $\bm{p} \in \mathbb{R}^{m \times 3}$ and define binary selector matrices $\mathbf{S}_i \in \mathbb{R}^{m \times m}$ such that $\bm{p}_i = \mathbf{S}_i \bm{p}$. The dimension $m$ is equal to $\sum_i c_i$ (summation over all quadratic energies), where $c_i$ is one for the bending energy, two for the local area energy and $n$ for the global volume conservation, i.e., it depends on the corresponding energy (see section \ref{sec:QuadraticEnergies}). More in details, the matrix $\mathbf{G}_i \in \mathbb{R}^{m \times n}$ formulates the deformation descriptor and places it in the involved rows and columns, where the columns correspond to the involved vertices of the $i$-th energy and the rows refer to the corresponding projection $\bm{p}_i$. The term ``constraint'' is used because the energy penalizes the deviation from the undeformed state, and thus \emph{constraints} the body from deformation, and the term ``projection'' corresponds to finding the minimum distance from the current state onto the manifold.

\begin{figure}[h]
    \centering
    \includegraphics[scale=0.4]{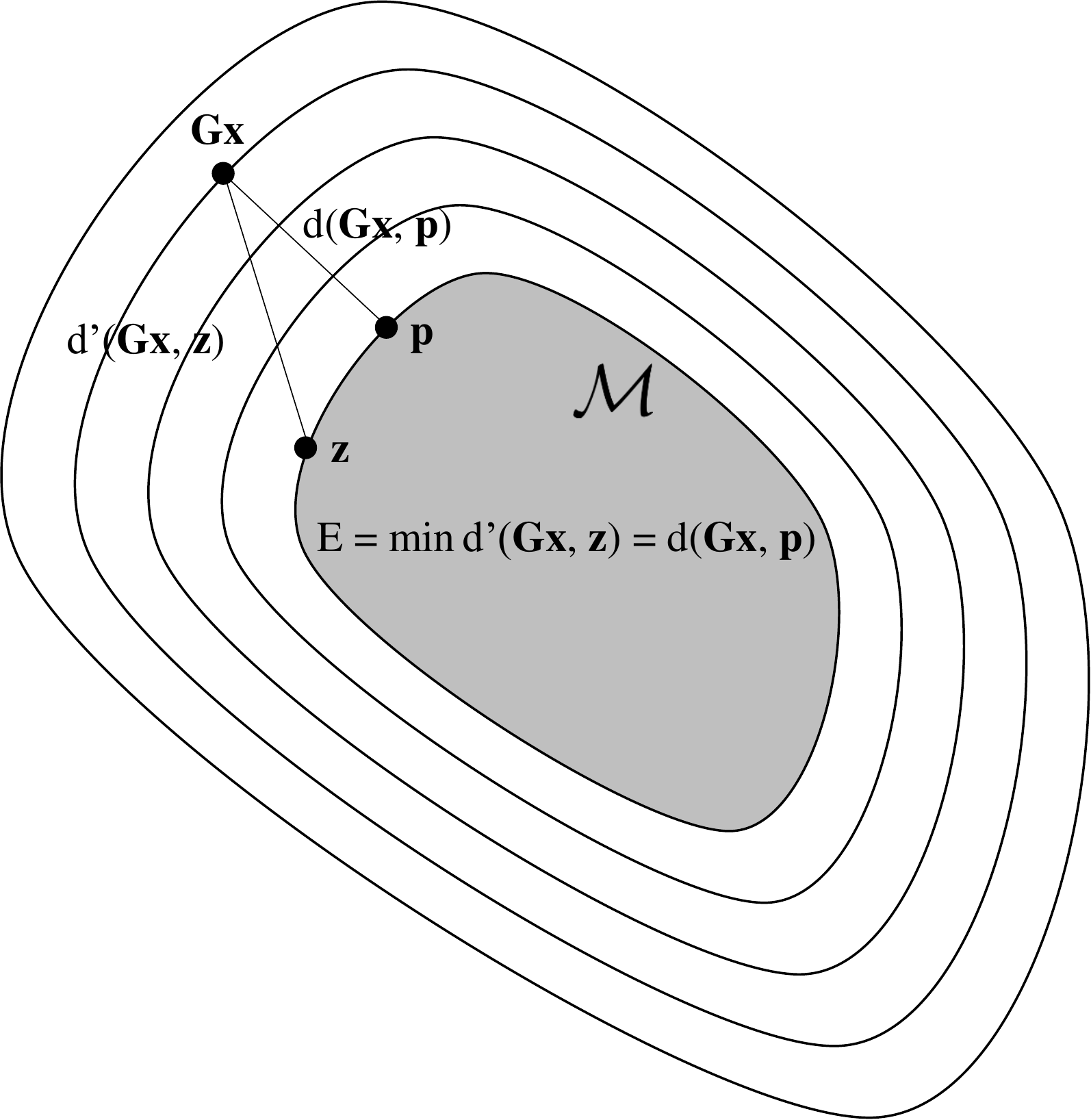}
    \caption{Geometric representation of a potential energy. The more the distance from the undeformed state $\mathcal{M}$ the higher the energy stored in the body. The body seeks to minimize its potential energy through $\mathbf{G}\bm{x} = \bm{p}$.}
    \label{fig:manifold}
\end{figure}

Generalizing the above, all the possible undeformed configurations define the so-called constraint manifold $\mathcal{M}$ (Cartesian product of the individual constraint manifolds $\mathcal{M}_i$), which mathematically is formulated as the zero level-set of the total potential energy. From a geometric point of view, the potential energy is a distance $d$ which quantifies how far the deformed state $\mathbf{G}\bm{x}$ is from the manifold $\mathcal{M}$ (rest state). The projection of the current state onto $\mathcal{M}$ is denoted by $\bm{p}$. There can be many user-defined manifolds. For example, a manifold for volume preservation does not guarantee area conservation and therefore, we have to define another manifold to guarantee the latter constraint. Recapitulating the above, the total potential energy can be formalized as
\begin{equation} \label{eq:DistanceFormulation}
    E(\bm{x}, \bm{p}) = \min_{\bm{z} \in \mathcal{M}} d^{'}(\mathbf{G}\bm{x}, \bm{z}) = \left \| \mathbf{G} \bm{x} - \bm{p} \right \|^{2}_{F}.
\end{equation}
The minimum is achieved when $\bm{p}$ is the projection of $\mathbf{G} \bm{x}$ onto $\mathcal{M}$. Figure \ref{fig:manifold} summarizes the definition of the distance and its correspondence to a potential energy. The generic potential energy of equation \eqref{eq:DistanceFormulation} refers to a discretized energy. Later, we show how from continuous energies using linear basis functions, we retrieve this generic formulation.

The key idea of projective dynamics is to exploit these specialized potential energies and reformulate the \emph{variational form of implicit Euler} as
\begin{equation} \label{eq:ObjectiveFunction}
    g(\bm{x}_{n+1}) = \frac{1}{2h^2} tr((\bm{x}_{n+1}-\bm{y}_n)^T \mathbf{\widetilde{M}} (\bm{x}_{n+1}-\bm{y}_n)) + \sum_i w_i \left ( \underset{\mathbf{S}_i \bm{p} \in \mathcal{M}_i}{\min} \widetilde{E}_i(\bm{x}, \mathbf{S}_i \bm{p}) \right ),
\end{equation}
where $g$ is the objective function whose minimum corresponds to $\bm{x}_{n+1}$ and $w_i = A_i k_i$ is a weighting term acting like a ``stiffness'' that converts the distance to an energy term ($A_i$ is the area of the element and $k_i$ is the stiffness factor). As a reminder, $||A||^{2}_{F} = tr(A^T A)$. The projections $\bm{p}$ are essentially functions of $\bm{x}$ and thus developing even further equation \eqref{eq:PDEnergy} into the variational form \eqref{eq:genericOptimizationNewton}, we get
\begin{equation} \label{eq:fullyDeveloped_g}
    \begin{split}
        g(\bm{x}_{n+1}) =& \frac{1}{2h^2} tr\left ( (\bm{x}_{n+1} - \bm{y}_n)^T \widetilde{\mathbf{M}} (\bm{x}_{n+1} - \bm{y}_n) \right ) + \\
        & \underbrace{\frac{1}{2} tr\left ( \bm{x}_{n+1}^T \mathbf{L} \bm{x}_{n+1} \right ) - tr\left ( \bm{x}^T_{n+1} \mathbf{J} \bm{p}(\bm{x}_{n+1}) \right ) + \frac{1}{2} tr\left ( \bm{p}^T(\bm{x}_{n+1})\mathbf{S}\bm{p}(\bm{x}_{n+1}) \right )}_{\text{projective dynamics potential energies: quadratic formulation}},
    \end{split}
\end{equation}
where $\mathbf{L} = \sum w_i \mathbf{G}_i^T \mathbf{G}_i$, $\mathbf{J} = \sum w_i \mathbf{G}_i^T \mathbf{S}_i$ and $\mathbf{S} = \sum w_i \mathbf{S}_i \mathbf{S}_i^T$.

Seeking for greater material expressivity, and thus more general potential energies, we can straightforwardly extend equation \eqref{eq:ObjectiveFunction} as
\begin{equation} \label{eq:GeneralizedObjectiveFunction}
    g(\bm{x}) = \frac{1}{2h^2} tr((\bm{x}-\bm{y})^T \mathbf{\widetilde{M}} (\bm{x}-\bm{y})) + \sum_i w_i E_i^{PD}(\bm{x}) + \sum_i E_i^{non-PD}(\bm{x}),
\end{equation}
where PD stands for energies that follow the quadratic formulation of projective dynamics, see equation \eqref{eq:DistanceFormulation}, and non-PD for arbitrary energies. A remark is the absence of a weighting term in the non-PD energies, which is because the stiffness is already integrated into them. The minimization of $g$ is performed using a quasi-Newton technique with a well-suited Hessian approximation, which is deeply rooted in the projective dynamics part. The first step is to compute the gradient of the objective function $g(\bm{x})$ as defined by equation \eqref{eq:GeneralizedObjectiveFunction}. An in-depth derivation of the gradient (regarding the PD part) and an explanation on why the derivative of $\frac{1}{2} tr\left ( \bm{p}^T(\bm{x}) \mathbf{S} \bm{p}(\bm{x}) \right )$ vanishes can be found in \citep{Liu2017Quasi-NewtonMaterials, Liu2018TowardsMaterials}. The derivative is given by
\begin{equation} \label{eq:GeneralizedObjectiveFunctionDerivative}
    \nabla g(\bm{x}) = \frac{1}{h^2} \widetilde{\mathbf{M}} \left (\bm{x} - \bm{y}  \right ) + \underbrace{\mathbf{L} \bm{x} - \mathbf{J} \bm{p}(\bm{x})}_{\mathclap{\text{$-\bm{F}_{int}^{PD}$}}} +
    \underbrace{\sum_i \nabla_{\bm{x}} E_i^{non-PD}(\bm{x})}_{\mathclap{\text{$-\bm{F}_{int}^{non-PD}$}}}.
\end{equation}

The projective dynamics energies demand first the computation of $\bm{p}$ for a given $\bm{x}$. This step can be massively parallelized given the small computational cost of the projection per element. The objective function could be equivalently minimized by Newton's method. Newton's method would proceed by computing the Hessian matrix $\nabla^2 g(\bm{x})$ and then using it to compute a descent direction as $d=-(\nabla^2 g(\bm{x}))^{-1} \nabla g(\bm{x})$. However, Liu et al. 2017 \citep{Liu2017Quasi-NewtonMaterials} suggested replacing the Hessian by $\widetilde{\mathbf{H}} = \widetilde{\mathbf{M}}/h^2 + \mathbf{L}$ and thus opting for a quasi-Newton approach avoiding the expensive computation of the Hessian at every time step and the possible definiteness fixes to guarantee that $d$ is a descent direction. Using only quadratic energies, the minimization of $g$ could be realized by an alternating (local/ global) solver \citep{Bouaziz2014ProjectiveSimulation}, which reveals how $\widetilde{\mathbf{H}}$ is formed. A detailed presentation can be found in \ref{appendix:LocalGlobal}. The approximated Hessian plays a major role in the fast convergence of our solver and in the overall computational efficiency of the npFEM solver without the expensive computation of the true Hessian. {\color{black} Furthermore, matrix $\widetilde{\mathbf{H}}$ is constant, symmetric positive definite and thus can be prefactorized once at the beginning, e.g., using Cholesky factorization, allowing for very fast computation of the descent direction $d=-(\widetilde{\mathbf{H}})^{-1} \nabla g(\bm{x})$.}

The non projective dynamic energies could contribute as well to the formulation of the Hessian approximation by considering their contribution to $\sum w_i \mathbf{G}_i^T \mathbf{G}_i$, where $\mathbf{G}_i$ is the deformation gradient operator for the $i$-th triangular element and $w_i = A_i k_i$ ($A_i$ is the area of the element and $k_i$ a stiffness factor which can be computed as described in \citep{Liu2017Quasi-NewtonMaterials}). If on the other hand there is no contribution from the non-PD part, then the iterations for convergence are slightly increased.

{\color{black} The overall convergence is affected by the selected potential energies. The more the PD energies, the faster the convergence, which is the case for our study since out of the four different potentials (area, volume, bending, material) only the material energy does not fall into the PD category, see sections \ref{sec:QuadraticEnergies} \& \ref{sec:correctedSkalak} for more details. The reason for this, is that the Hessian approximation is mainly built upon the PD part.}

It is worth noting that the size of our matrices is $n \times n$ instead of $3n \times 3n$, typically found in other FEM solvers \citep{Macmeccan2009SimulatingMethod}, leading to an at least 3$\times$ speedup compared to them.

Consequently, moving towards the descent direction, this iterative procedure guarantees to converge to the minimum of $g$, i.e., $\bm{x}_{n+1}$. A recapitulation of the algorithm is presented in section \ref{sec:npFEMRecap}.

\subsubsection{Quadratic Energies to simulate RBC membrane} \label{sec:QuadraticEnergies}
In this section, we develop potential energies suitable for projective dynamics and critical for the simulation of the red blood cell membrane, namely local area conservation, bending and global volume preservation. The analysis reveals that the above-mentioned energies can be written in the quadratic form described by equation \eqref{eq:PDEnergy}. The goal is to develop discrete energies and compute the auxiliary variables, i.e., the projections, in order to quantify the energy term itself. Since the current state $\bm{x}$ is known, the computation of $\bm{p}$ results in the quantification of the energy $E$ from equation \eqref{eq:PDEnergy}.

Let the undeformed surface be a differentiable 2-manifold surface $S$ embedded in $\mathbb{R}^3$. Let us define piecewise linear coordinate functions of the undeformed and deformed surface, $\bm{g}: S \rightarrow \mathbb{R}^3$ and $\bm{f}: S \rightarrow \mathbb{R}^3$, respectively. By introducing a set $M$ of desired point-wise transformations $\mathbf{T}$, we can formulate a potential energy \citep{Bouaziz2014ProjectiveSimulation} as
\begin{equation} \label{eq:GenericContinuousPotential}
    E( \bm{f}, \mathbf{T}) = \frac{k}{2} \int_{S} \left \| \nabla_S \bm{f} - \mathbf{T} \nabla_S \bm{g} \right \|_F^2 dA,
\end{equation}
where $\nabla_S$ is the gradient operator on $S$. The set $M$ determines all allowed rest configurations $\mathbf{T} \nabla_S \bm{g}$. The energy $E\left ( \bm{f}, \mathbf{T} \right )$ is discretized over triangular elements using piecewise linear hat functions \citep{Botsch2010PolygonProcessing}, and the integral is transformed to a sum over triangle potentials:
\begin{equation}
    E_i(\bm{x}, \mathbf{T}) = \frac{k_i}{2} A_i || \mathbf{D_s} \mathbf{D_m}^{-1} - \mathbf{T} ||_F^2,
\end{equation}
where $A_i$ is the triangle area, $\mathbf{D_s} = [\bm{x}_j - \bm{x}_i, \bm{x}_k - \bm{x}_i] \in \mathbb{R}^{2 \times 2}$ contains the triangle edges of the deformed state embedded in 2D and $\mathbf{D_m}$ corresponds to the undeformed state. At this point, we have introduced a generic quadratic energy similar to what is described by equation \eqref{eq:PDEnergy}. Based on the FEM reminder of \ref{appendix:FEMReminder}, the generic quadratic energy defined above can be rewritten as
\begin{equation} \label{eq:discretizedGenericEnergy}
    E_i(\bm{x}, \mathbf{T}) = \frac{k_i}{2} A_i || \mathbf{F} - \mathbf{T} ||_F^2,
\end{equation}
where $\mathbf{F} \in \mathbb{R}^{2 \times 2}$ is the deformation gradient of the $i$-th triangular element. In this case, the deformation gradient acts like a deformation descriptor, and $\mathbf{T}$ corresponds to the projection on the manifold defined by the set $M$.

\textbf{Area preservation energy.} Let us define the set $M$ as the set of matrices with determinant 1, i.e., $det(\mathbf{T}) = 1$. In mathematical terms, this requirement is equivalent to $M \equiv SL(2)$, where the special linear group of degree n, $SL(n)$, is the set of $n \times n$ matrices with determinant 1. Let us consider the singular value decomposition (SVD) of the deformation gradient of a triangular element
\begin{equation}
    \mathbf{F} = \mathbf{U} \mathbf{\Sigma} \mathbf{V}^T, 
\end{equation}
where $\mathbf{\Sigma}$ is a diagonal matrix storing the singular values or \emph{principal stretches} $\lambda_1, \lambda_2$ of $\mathbf{F}$. Area conservation of the element is equivalent to
\begin{equation}
    det(\mathbf{F}) = det(\mathbf{\Sigma}) = 1.
\end{equation}
Consequently, the generic quadratic energy is rewritten as
\begin{equation}
    \frac{k_i}{2} A_i || \mathbf{\Sigma} - \mathbf{\Sigma}^* ||_F^2 ~~\text{s.t.}~~\text{det}(\mathbf{\Sigma}^*) = 1,
\end{equation}
which is the distance of the current state $\mathbf{\Sigma}$ from the manifold defined by $SL(2)$. A rigorous solution to this projection $\mathbf{\Sigma}^*$ can be found in the Appendix of \citep{Bouaziz2014ProjectiveSimulation}. Essentially, this energy penalizes how far the element is from the state that conserves the area $\mathbf{\Sigma}^*$.

\textbf{Bending Energy.} The RBC membrane simulation demands a bending energy, commonly formulated by employing the dihedral angles across edges. The bending energy used in this study \citep{Bouaziz2014ProjectiveSimulation} measures the squared difference of mean curvatures $H$ as
\begin{equation}
    E(\bm{f}, \mathbf{R}) = \frac{k}{2} \int_S || \bm{\Delta}_S\bm{f} - \mathbf{R} \bm{\Delta}_S \bm{g} ||_2^2~dA,
\end{equation}
where $\mathbf{R} \in SO(3)$ are rotation matrices that best match the initial to the deformed state, $\bm{\Delta}_S$ is the Laplace-Beltrami operator on $S$. As a reminder, this operator applied on the coordinate function $\{\bm{g},\bm{f}\}$ of the surface, gives $\bm{\Delta}_S \{\bm{g},\bm{f}\} = -2H_{\{\bm{g},\bm{f}\}}\bm{n}$, where $H$ is the mean curvature and $\bm{n}$ is the surface normal. The discretized version \citep{Bouaziz2014ProjectiveSimulation} of the bending energy per vertex $i$, using piecewise linear hat functions, is
\begin{equation} \label{eq:BendingDiscretizedPotential}
    E_i \left ( \bm{x}, \mathbf{R} \right ) = \frac{k_i}{2} A_i \left \| \bm{X}_f \bm{c} - \mathbf{R} \bm{X}_g \bm{c} \right \|_2^2,
\end{equation}
where $A_i$ is the Voronoi area of the vertex $i$, $\bm{X}_f$ and $\bm{X}_g$ contain the one-ring edges of the vertex in the current and initial configuration, respectively, and $c$ stores the common cotangent weights divided by the Voronoi area. The projection $\mathbf{R}$ has a closed form solution and is presented in the Appendix of \citep{Bouaziz2014ProjectiveSimulation}. For a thorough analysis on the different bending energy algorithms, consider the work of Guckenberger et al. 2016 \citep{Guckenberger2016OnFlows}.

\textbf{Global volume conservation.} The volume $V$ of a body embedded in 3D can be computed employing the divergence theorem as
\begin{equation}
    V = \frac{1}{3} \underset{\mathcal{V}}{\iiint} \nabla \cdot \bm{x} d\bm{x} = \frac{1}{3} \underset{\mathcal{\partial V}}{\iint} \bm{x}^T \bm{n} d\bm{x},
\end{equation}
where $\mathcal{\partial V}$ is the boundary of the 3D body $\mathcal{V}$ (RBC membrane in our case) and $\bm{n}$ is the surface normal. Given the tessellation of the membrane in triangular elements, we can further simplify \citep{Bender2017A2017} the volume as
\begin{equation}
    V(\bm{x}) = \frac{1}{3} \underset{\mathcal{\partial V}}{\iint} \bm{x}^T \bm{n} d\bm{x} = \frac{1}{9} \sum_i A_i (\bm{x}_{i_1} + \bm{x}_{i_2} + \bm{x}_{i_3})^T \bm{n}_i,
\end{equation}
where $A_i$ is the area and $i_1, i_2, i_3$ are the vertex indices of the $i$-th element. Having defined the above toolset, we can now formulate the global volume conservation manifold as $C = V(\bm{x}) - V_0 = 0$, where $V_0$ is the volume at rest. Therefore, the global volume conservation energy can be written as
\begin{equation}
    E_i(\bm{x}, \bm{p}) = \frac{w_i}{2} \left \| \bm{p} - \bm{x} \right \|_F^2~~\text{s.t.}~~\bm{p} \in C.
\end{equation}
The deformation descriptor is the absolute positions of the deformed body, and the projection corresponds to a state vector which respects the volume conservation.

To find the projection $\bm{p}$, we reformulate the above energy into a minimization problem as
\begin{equation}
\begin{split}
    & \min_{\bm{p}}~~\frac{w_i}{2} \left \| \bm{p} - \bm{x} \right \|_F^2, \\ & \text{s.t.}~~C(\bm{p}) = C(\bm{x} + \Delta \bm{x}) = 0 \\
    & ~~~~~~\Delta \bm{x} = \lambda \nabla C(\bm{x}),
\end{split}
\end{equation}
the second constraint that restricts $\Delta \bm{x}$ to be in the direction of $\nabla C$ is a requirement for linear and angular momentum conservation \citep{Bender2017A2017}. The solution to the above optimization problem using Lagrange multipliers is \citep{Bender2017A2017, Bouaziz2014ProjectiveSimulation}
\begin{equation}
    \Delta \bm{x} = - \frac{C(\bm{x})}{\left \| \nabla C(\bm{x}) \right \|_F^2} \nabla C(\bm{x}),
\end{equation}
while the gradient can be approximated \citep{Bender2017A2017} by
\begin{equation}
    \nabla C(\bm{x}) \approx \frac{1}{3} [ \bar{\bm{n}}_1^T, ..., \bar{\bm{n}}_n^T ]^T,
\end{equation}
where $\bar{\bm{n}}_k = \sum A_j \bm{n}_j$ is the sum of the area weighted normals of all elements which contain vertex $k$.

Global area conservation could be imposed by a similar approach. Nevertheless, after extensive testing we found out that only the local area preservation term is enough to describe the RBC behavior, without increasing more the computational cost. Essentially, since the area preservation is a property of the material, it is sufficient to encode it locally, through the elements that compose the body.

The energies developed in this section follow the quadratic formulation described by equation \eqref{eq:PDEnergy}. However, to build the stacked system of projections as described by equation \eqref{eq:fullyDeveloped_g}, the user has to place carefully each deformation descriptor and projection to the sparse matrices $\mathbf{G}_i \bm{x}$ and $\bm{p}_i$. Additionally, the descriptors and projections isometrically embedded in $\mathbb{R}^2$, have to be re-embedded back in $\mathbb{R}^3$, since the equations of motion refer to three dimensions.

\subsubsection{Non-PD energies \& internal forces} \label{sec:correctedSkalak}
There are many energies following the above-mentioned quadratic distance formulation but the simulated materials are quite limited. Even classical models from continuum mechanics, such as Neo-Hookean, St. Venant-Kirchhoff, Mooney-Rivlin, are not supported. More importantly, the RBC falls into the latter category and for this reason, we develop the non projective dynamic energies and forces. 

Consequence of elastic deformation is the accumulation of potential energy in the body. Our focus is on hyperelastic materials for which the potential energy $E[\bm{\phi}]$ is determined by the deformation function $\bm{\phi}$ (see \ref{appendix:FEMReminder}) and is independent of the deformation history. To describe a deformed body, we introduce the energy density function $\Psi[\bm{\phi}; \bm{X}]$ which measures the potential energy per unit undeformed volume/ area on an infinitesimal domain $dV$ around the material point $\bm{X}$. The total potential energy is
\begin{equation}
    E[\phi] = \int_{V} \Psi [\bm{\phi}; \bm{X}] d\bm{X} = \int_{V} \Psi [\mathbf{F}] d\bm{X}.
\end{equation}

Every material follows a specific constitutive model which links the stresses with the strains. There is a variety of stress descriptors; for our discussion we use the $1^{st}$ Piola-Kirchhoff stress tensor $\mathbf{P}$ which for hyperelastic materials is
\begin{equation}
    \mathbf{P}(\mathbf{F}) = \partial \Psi(\mathbf{F}) / \partial \mathbf{F}.
\end{equation}

The assumption of isotropy and rotation invariance allows us to express the energy density function $\Psi$ as a function of the principal stretches of the deformation gradient ($\lambda_1, \lambda_2$ for triangular elements). To simplify the material space, we use the \emph{Valanis-Landel hypothesis} \citep{Valanis1967TheRatios} which assumes that
\begin{equation}
    \Psi(\lambda_1, \lambda_2) = f(\lambda_1) + f(\lambda_2) + g(\lambda_1 \lambda_2),
\end{equation}
where $f, g$ are scalar non-linear elastic strain energy density functions. The original formulation of the hypothesis is for solid bodies (typically rubber-like materials) and not for surfaces, thus involving the three principal stretches ($\lambda_1, \lambda_2, \lambda_3$) of the deformation gradient and three scalar functions ($f,g,h$). It covers a vast amount of materials, such as St. Venant-Kirchhoff, Neo-Hookean and Mooney-Rivlin.

The well-known energy introduced by Skalak et al. 1973 \citep{Skalak1973StrainMembranes} for the description of the membrane of a red blood cell falls into the category of the Valanis-Landel materials. Given $I_1 = \lambda_1^2 + \lambda_2^2 - 2$ and $I_2 = \lambda_1^2 \lambda_2^2 - 1$ the energy density function is
\begin{equation}
    \Psi = \frac{B}{4} \left (\frac{1}{2} I_1^2 + I_1 - I_2 \right) + \frac{C}{8}I_2^2,
\end{equation}
where $B$ and $C$ are membrane material properties. The above energy density function can be alternatively written in the Valanis-Landel form as
\begin{align}
    f(\lambda_i) &= \frac{B}{8} \lambda_i^4 - \frac{B}{4} \lambda_i^2, \\
    g(\lambda_i \lambda_j) &= \frac{C}{8} (\lambda_i \lambda_j)^4 - \frac{C}{4} (\lambda_i \lambda_j)^2.
\end{align}
Given that we enforce area conservation through the projective dynamics part, i.e., $\lambda_1 \lambda_2 = 1$, $g$ can be neglected and only $f$ should be considered.

Since we are dealing with linear finite elements, the deformation gradient is constant over each element ($\mathcal{T}_i$), see \ref{appendix:FEMReminder}, and this observation reduces the elemental potential energy to
\begin{equation}
    E_i = \int_{\mathcal{T}_i} \Psi(\mathbf{F}) d\bm{X} = W_i \cdot \Psi(\mathbf{F}_i),
\end{equation}
where $W_i$ is the undeformed area of the element. The contribution of element $\mathcal{T}_i$ to the elastic forces is $\bm{F}_{int}^{non-PD} = - \nabla E_i$ which can be further simplified for a triangular element to \citep{Sifakis:2012:FSD:2343483.2343501} 
\begin{equation}
    [\vec{f_1}~~\vec{f_2}] = - W_i \mathbf{P}(\mathbf{F}) \mathbf{D_m}^{-T}~~\text{and}~~ \vec{f_3} = -\vec{f_1}-\vec{f_2}, 
\end{equation}
where $\vec{f_1},\vec{f_2},\vec{f_3}$ are the internal forces on the vertices of $\mathcal{T}_i$. Moreover, the $1^{st}$ Piola-Kirchhoff stress tensor can be further simplified to \citep{Xu2015NonlinearStretches}
\begin{equation}
    \mathbf{P}(\mathbf{F}) = \mathbf{P}(\mathbf{U} \mathbf{\Sigma} \mathbf{V}^T) = \mathbf{U} \mathbf{P}(\mathbf{\Sigma}) \mathbf{V}^T,
\end{equation}
where $\mathbf{P}(\mathbf{\Sigma}) = \text{diag} \left ( \frac{\partial \Psi}{\partial \lambda_1}, \frac{\partial \Psi}{\partial \lambda_2} \right )$.

\textbf{Stability.} Special care is needed when designing new materials to satisfy Drucker's first stability criterion (also known as Hill's stability criterion), which requires a monotonic increase of strain energy density with increase in strain \citep{Drucker1957AMaterial}. In practice, this stability criterion is satisfied by demanding $f'', g'' > 0$ \citep{Xu2015NonlinearStretches}.

\textbf{Modification on Skalak's Energy.} This energy behaves quite well for small deformations as we present in section \ref{sec:Results}, but for larger deformations it fails to robustly describe the mechanics of the RBC membrane. For this reason, we enhance the strain hardening behavior by adding one more term in $f$. The modified version is as follows:
\begin{equation}
    f(\lambda_i) = \frac{B}{8} \lambda_i^4 - \frac{B}{4} \lambda_i^2 + \frac{D}{4} (\lambda_i - 1)^4
\end{equation}
where $D$ is a material property like $B$. The modified Skalak's energy satisfies Drucker's stability criterion and has a superior behavior than the original one, see section \ref{sec:Results}. The correction term has zero contribution at rest, where $\lambda_i = 1$, and its form was defined through fitting on the available experimental data.

\subsubsection{Membrane viscoelasticity} \label{sec:Viscoelasticity}
Membrane viscoelasticity is tackled in a two-way approach. Initially, through the addition of the Rayleigh dissipation term in Newton's equation \eqref{eq:RayleighNewton} and secondly through a post-convergence step.

Typically, the Rayleigh damping matrix is a blend of the mass and stiffness matrices
\begin{equation}
    C = \alpha_D \mathbf{M} + \beta_D \mathbf{K},
\end{equation}
where $\mathbf{M} \in \mathbb{R}^{n \times n}$ is the mass matrix, $\mathbf{K} \in \mathbb{R}^{n \times n}$ is the stiffness matrix and $\alpha_D, \beta_D$ are coefficients. In our analysis we do not construct explicitly the stiffness matrix (Hessian of the elastic potential energy), and thus we are bypassing this issue by using matrix $\mathbf{L} = \sum w_i \mathbf{G}_i^T \mathbf{G}_i$ instead \citep{li2018laplacian}. Essentially, we make use of a part of the Hessian approximation instead of the Hessian itself. A critical point to consider is that Rayleigh damping can affect the rigid body motions (translation, rotation) if the damping ratio is high. The damping ratio ($\zeta$) is defined as $\zeta(\omega_n) = 0.5(\alpha_D \omega_n^{-1} + \beta_D \omega_n)$, where $\omega_n$ is the $n$-th modal frequency \citep{Macmeccan2009SimulatingMethod}. In our study, we do an approximate modal analysis by solving $\mathbf{L} \bm{x} = \bm{\omega}^2 \mathbf{M} \bm{x}$. For our analysis, we set $\alpha_D = 0$ and make sure that $\zeta(\omega_n) \ll 1$ for every mode. {\color{black}In this way, we guarantee that the trajectory of the simulated bodies follows the proper dynamics and that the momenta are conserved.}

Due to the above restriction of Rayleigh damping, we cannot fully emulate the viscoelastic behavior of the RBC membrane. {\color{black}This is why we introduce the post-convergence step that complements Rayleigh damping, does not affect the rigid body dynamics, and also conserves linear \& angular momenta.} The key idea is to calculate the difference between the vertex velocity $\bm{v}_i$ and its velocity in the best-fit rigid body motion. Then, we dissipate partially the non-rigid-body part. The algorithm \citep{Muller2007PositionDynamics} to do so is summarized in the following lines:
\begin{algorithmic}
\State $\bm{x}_{cm} = (\sum_i \bm{x}_i m_i) / (\sum_i m_i)$ \Comment{center of mass (cm)}
\State $\bm{v}_{cm} = (\sum_i \bm{v}_i m_i) / (\sum_i m_i)$ \Comment{velocity of cm}
\State $\mathbf{L} = \sum_i \bm{r}_i \times (m_i \bm{v}_i)$ \Comment{angular momentum}
\State $\mathbf{I} = \sum_i \widetilde{\bm{r}}_i \widetilde{\bm{r}}_i^T m_i$ \Comment{inertia matrix}
\State $\bm{\omega} = \mathbf{I}^{-1} \mathbf{L}$ \Comment{angular velocity}
\ForAll{vertices~~i} 
\State $\Delta \bm{v}_i = \bm{v}_{cm} + \bm{\omega} \times \bm{r}_i - \bm{v}_i$
\State $\bm{v}_i \gets \bm{v}_i + \kappa_{\text{damping}} \Delta \bm{v}_i$
\EndFor
\end{algorithmic}
where $\bm{r}_i = \bm{x}_i - \bm{x}_{cm}$, $\widetilde{\bm{r}}_i \in \mathbb{R}^{3 \times 3}$ is a matrix with the property $\widetilde{\bm{r}}_i \bm{v} = \bm{r}_i \times \bm{v}$, and $\kappa_{\text{damping}} \in [0,1]$ is the damping coefficient. This dissipation term is deployed after the convergence to $\bm{x}_{n+1}$ (see algorithm \ref{algo:npFEM}).

An important remark to avoid confusion is that in our study we are not simulating a viscoelastic material, i.e., the potential energy to be dependent on the strain rate, but we are emulating the viscoelastic behavior through viscous dissipation (proxy for viscoelasticity). This approach is proven to have equivalent rheological behavior with the \emph{generalized Maxwell model} as proposed by Semblat 1997 \citep{Semblat1997RheologicalDamping}. Similar approach is followed by Fedosov et al. 2010 \citep{Fedosov2010ARheologydynamics} \& Hemocell \citep{Zavodszky2017CellularCells}.

\subsubsection{Deformable bodies solver: npFEM algorithm} \label{sec:npFEMRecap}
In this section, we summarize the algorithm for the deformable bodies, focused on the simulation of red blood cells (algorithm \ref{algo:npFEM} describes one time step of the npFEM solver). Its extension to any other type of cell or body is based on the same principles. The advancement of the body from time $t$ to $t+1$ is done through the minimization of equation \eqref{eq:GeneralizedObjectiveFunction}. The main constituents of this equation are two categories of energies. The former follows the projective dynamics setup with quadratic energies. The latter are energies that do not belong in any particular category, but they offer a more advanced description of the material space.

Equation \eqref{eq:GeneralizedObjectiveFunction} is a typical non-linear unconstrained minimization problem for which we employ a quasi-Newton technique. To further speed-up the convergence, we deploy the LBFGS algorithm \citep{Nocedal2006NumericalOptimization} that takes into consideration previous iterates and gives a better Hessian approximation based on the initial one. {\color{black} Moreover, given the positive definiteness of $\widetilde{\mathbf{H}}$, $\bm{d}(\bm{x})$ is guaranteed to be a descent direction. However, to guarantee a sufficient and optimal decrease, we implement a popular inexact line search technique based on \emph{Armijo} and \emph{curvature} conditions, collectively known as \emph{Wolfe conditions}. For a detailed overview on quasi-Newton techniques consider the seminal work of Nocedal \& Wright \citep{Nocedal2006NumericalOptimization}.}

{\color{black}
Regarding the interior fluid of the red blood cell (cytoplasm), we simulate it implicitly by considering the global volume conservation (incompressibility of cytoplasm) and through the viscoelasticity of the full body as described in section \ref{sec:Viscoelasticity}. The decision not to simulate the interior fluid of RBCs is not because of some limitation of the computational framework but rather a matter of computational efficiency. We could have simulated the interior fluid through Palabos and subsequently to tune the viscoelastic terms to correspond to the membrane only. In this case, the global volume conservation energy has to be removed, since it would be implicitly conserved through the fluid solver, while the rest of the energies (area, bending, material) do not have to change. Moreover, from the literature there is no indication that the simulation of the interior fluid plays a crucial role on accurately reproducing the collective transport of blood cells, and for this reason we have decided to follow the more lightweight implementation path.
}

\begin{algorithm}[h]
\caption{npFEM solver: The algorithm takes the deformable body from time $t \rightarrow t+1$}
\label{algo:npFEM}
\begin{algorithmic}[1]
\State $\bm{x}^{(1)} \gets \bm{x}_{n} + h \widetilde{\mathbf{M}}^{-1} \mathbf{M} \bm{v}_n + h^2 \widetilde{\mathbf{M}}^{-1} \bm{F}_{ext} $ \Comment{First guess}
\While{convergence, $k$++}
\State $\bm{d} \left ( \bm{x}^{(k)} \right ) \gets \text{LBFGS} \left ( \widetilde{\mathbf{H}} \right )$ \Comment{Descent direction: Algorithm 7.4 \citep{Nocedal2006NumericalOptimization}}
\State $\alpha \gets 2$
\Repeat \Comment{Line search}
\State $\alpha \gets \alpha / 2$
\State $\bm{x}^{(k+1)} \gets \bm{x}^{(k)} + \alpha \bm{d} \left ( \bm{x}^{(k)} \right )$
\State evalProjectionsAndInternalForces$\left ( \bm{x}^{(k+1)} \right )$ \Comment{PD \& non-PD part}
\State $g \left ( \bm{x}^{(k+1)} \right ) \gets \text{evalObjective}()$
\State $\nabla g \left ( \bm{x}^{(k+1)} \right ) \gets \text{evalGradient}()$
\Until{Wolfe conditions}
\EndWhile \Comment{Converged to $\bm{x}_{n+1}$}
\State $\bm{v}_{n+1} \gets (\bm{x}_{n+1} - \bm{x}_{n}) / h$ 
\State DissipationTerm\_PostConvergence() \Comment{section \ref{sec:Viscoelasticity}}
\end{algorithmic}
\end{algorithm}

\subsection{Lattice Boltzmann Method (LBM)} \label{sec:LBM}
The lattice Boltzmann method (LBM) is employed to solve indirectly the Navier-Stokes equations. Densities of virtual particles, also known as populations, move on a regular grid/ lattice and collide at the lattice nodes. The LBM originally grew out of the lattice gas models \citep{Hardy1973TimeFunctions, Frisch1986Lattice-GasEquation} and while lattice gases track particles, LBM tracks their statistical distribution. LBM is a second-order accurate solver for the weakly compressible Navier-Stokes equation, where the weak compressibility refers to errors that amplify as Mach number tends to one \citep{Kruger2017TheMethod}.

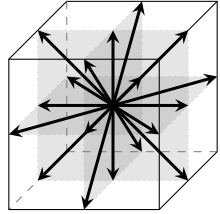
\begin{figure}[h]
    \centering
    \begin{tikzpicture}[scale=1.0,baseline={(current bounding box.center)}]
    \draw (2,0,0)--(2,2,0)--(0,2,0); 
    \draw [gray,dashed] (0,0,0)--(2,0,0);
    \draw [gray,dashed] (0,0,0)--(0,2,0);
    \draw [gray,dashed] (0,0,0)--(0,0,2);
    \draw (2,0,0) -- (2,0,2);
    \draw (0,2,0) -- (0,2,2);
    \draw [dotted,fill=gray, opacity=0.2] (1,0,0)--(1,2,0)--(1,2,2)--(1,0,2)--cycle; 
    \draw [dotted,fill=gray, opacity=0.2] (0,0,1)--(0,2,1)--(2,2,1)--(2,0,1)--cycle; 
    \draw [dotted,fill=gray, opacity=0.2] (0,1,0)--(0,1,2)--(2,1,2)--(2,1,0)--cycle; 
    \begin{scope}[shift={(1,1,1)}]
    \draw (0.003,-0.01) node {$\bullet$};
    \draw [line width=0.4mm,>=stealth, ->] (0,0,0) -- (1,1,0);
    \draw [line width=0.4mm,>=stealth, ->] (0,0,0) -- (-1,1,0);
    \draw [line width=0.4mm,>=stealth, ->] (0,0,0) -- (-1,-1,0);
    \draw [line width=0.4mm,>=stealth, ->] (0,0,0) -- (1,-1,0);
    \draw [line width=0.4mm,,>=stealth, ->] (0,0,0) -- (1,0,1);
    \draw [line width=0.4mm,>=stealth, ->] (0,0,0) -- (0,1,1);
    \draw [line width=0.4mm,>=stealth, ->] (0,0,0) -- (-1,0,1);
    \draw [line width=0.4mm,>=stealth, ->] (0,0,0) -- (0,-1,1);
    \draw [line width=0.4mm,,>=stealth, ->] (0,0,0) -- (1,0,-1);
    \draw [line width=0.4mm,>=stealth, ->] (0,0,0) -- (0,1,-1);
    \draw [line width=0.4mm,>=stealth, ->] (0,0,0) -- (-1,0,-1);
    \draw [line width=0.4mm,>=stealth, ->] (0,0,0) -- (0,-1,-1);
    \draw [line width=0.4mm,,>=stealth, ->] (0,0,0) -- (1,0,0);
    \draw [line width=0.4mm,>=stealth, ->] (0,0,0) -- (0,1,0);
    \draw [line width=0.4mm,>=stealth, ->] (0,0,0) -- (0,0,1);
    \draw [line width=0.4mm,>=stealth, ->] (0,0,0) -- (-1,0,0);
    \draw [line width=0.4mm,>=stealth, ->] (0,0,0) -- (0,-1,0);
    \draw [line width=0.4mm,>=stealth, ->] (0,0,0) -- (0,0,-1);
    \end{scope}
    \draw (0,0,2)--(2,0,2)--(2,2,2)--(0,2,2)--cycle; 
    \draw (2,2,0) -- (2,2,2);
    \end{tikzpicture}
    \caption{Sketch of the D3Q19 lattice structure on a lattice node.}
    \label{fig:D3Q19}
\end{figure}

The basic quantity of the LBM is the discrete-velocity distribution function $f_i(\bm{x},t)$, also known as particle populations. It represents the density of particles with velocity $\left \{ \bm{c}_i  \right \}_{i=0}^{q-1}$ at position $\bm{x}$ and time $t$. The number of populations $q$ per lattice node defines the resolution of the velocity space discretization. The points $\bm{x}$ are equally spaced in a regular 3D grid with a lattice spacing at $\Delta x$. By discretising the Boltzmann equation (which describes the evolution of the continuous distribution function $f$) in velocity space, physical space and time, we get the lattice Boltzmann equation (LBE)
\begin{align} \label{eq:LBE}
    f_i(\bm{x}, t) \gets f_i(\bm{x}, t) + \Omega_i(\bm{x}, t)~~~~~&(\text{Collision})\\
    f_i(\bm{x}+\bm{c}_i\Delta t, t+\Delta t) = f_i(\bm{x}, t)~~~~~&(\text{Streaming}),
\end{align}
which describes the collision and propagation of the mesoscopic particle packets. The velocities $\left \{ \bm{c}_i  \right \}_{i=0}^{q-1}$ are specially chosen such that $\bm{c}_i \Delta t$ point to a neighboring lattice site, with the consequence that populations are never trapped between nodes. For the collision operator $\Omega$, we are using the so-called Bhatnagar-Gross-Krook (BGK) approach \citep{Bhatnagar1954ASystems} 
\begin{equation}
    \Omega_i(f) = - \frac{f_i - f_i^{eq}}{\tau},
\end{equation}
which relaxes the populations towards an equilibrium $f^{eq}$ at a rate determined by the relaxation time $\tau$. The dimensionless relaxation parameter $\tau$ is linked to the speed of sound $c_s$ and the physical kinematic viscosity $\nu$ by $\nu = c_s^2 (\tau - 1/2) \Delta t$, where $c_s = \sqrt{1/3} \Delta x / \Delta t$ ($\Delta x, \Delta t$ space and time discretization, respectively). The equilibrium populations are given by
\begin{equation}
    f_i^{eq} = w_i \rho \left ( 1 + 3\bm{c}_i\cdot\bm{u} + \frac{9}{2}(\bm{c}_i\cdot\bm{u})^2 - \frac{3}{2} \bm{u}\cdot\bm{u} \right ),
\end{equation}
where $w_i$ are lattice weights and $\bm{u}$ is the macroscopic velocity at $\bm{x}$ and time $t$. In the present study, we use a 3D model with $q=19$ velocities, namely D3Q19. The lattice structure, the velocities $c_i$ and the lattice weights $w_i$ are thoroughly described in Kr{\"{u}}ger and colleagues 2017 \citep{Kruger2017TheMethod}. A sketch of the D3Q19 lattice is shown in Figure \ref{fig:D3Q19}.

The macroscopic properties of the fluid such as the density $\rho$, the velocity $\bm{u}$ and the hydrodynamic stress tensor $\bm{\sigma}$ can directly be recovered at each lattice node per time step as follows
\begin{align}
    \rho = \sum_i f_i \label{eq:LBM_dens}, \\
    \rho \bm{u} = \sum_i \bm{c}_i f_i \label{eq:LBM_Velocity}, \\
    \bm{\sigma} = -p\mathbf{I} + \left ( \frac{1}{2\tau} - 1\right )\sum_i \bm{c}_i\bm{c}_i f_i^{neq} \label{eq:stressTensor},
\end{align}
where $f_i^{neq} = f_i - f_i^{eq}$.

To further account for the presence of immersed objects, a forcing term $\bm{f}_{imm}$ should be added to the system. The method used in the present study is proposed in Shan \& Chen 1993 \citep{Shan1993LatticeComponents} and applies a correction to the momentum term of the equilibrium populations. The method is summarized in algorithm \ref{algo:LBM}.

\begin{algorithm}[h]
\caption{Lattice Boltzmann Method with forcing term (Shan \& Chen approach \citep{Shan1993LatticeComponents})}
\label{algo:LBM}
\begin{algorithmic}[1]
\State Compute macroscopic properties: $\rho, \bm{u}$
\State Compute $f_{imm}$ as described in section \ref{sec:IBM}
\State Corrected velocity: $\bm{u}^{G} = \bm{u} + \tau \Delta t \bm{f}_{imm}$
\State $\left \{ f_i^{eq}(\rho, \bm{u}^{G}) \right \}_{i=0}^{q-1}, q=19$
\State Collision step
\State Streaming Step ($t \rightarrow t+1$)
\end{algorithmic}
\end{algorithm}


\subsection{Immersed Boundary Method (IBM)} \label{sec:IBM}
The coupling between the red blood cells membrane and the blood plasma is realized through the immersed boundary method (IBM). Essentially, the IBM imposes a no-slip boundary condition, so that each point of the surface and the ambient fluid move with the same velocity. The original version of this method was developed by Peskin \citep{Peskin1972FlowMethod, Peskin2002TheMethod} to study heart valves. The key idea is that the surface of the deformable object is viewed as a set of Lagrangian points, which do not have to conform with the fluid mesh/ lattice. The fluid feels the presence of the body only via a force field which is added to the momentum conservation equations (see algorithm \ref{algo:LBM}). The interaction between the off-lattice marker points and the fluid is done via interpolation stencils, i.e., regularized Dirac delta functions. There are many variations of the IBM \citep{Mittal2005ImmersedMethods} and in this study we are using a modified version of the multi-direct forcing method introduced by Wang and colleagues 2008 \citep{Wang2008CombinedParticles}, in a form close to the one proposed by Ota et al. 2012 \citep{Ota2012LiftSimulations}.

Our goal is to compute a forcing term $f_{imm}$ along the immersed boundary and apply it as body force to the fluid. Let us denote with $\bm{x}$ and $\bm{X}_k$ the Eulerian and Lagrangian points respectively, with $\bm{U}_k$ the velocity of the Lagrangian point $k$ as computed by the solid solver (see algorithm \ref{algo:npFEM}) and with $\bm{u}^*(\bm{x}, t)$ the fluid velocity as computed by equation \eqref{eq:LBM_Velocity}. Using the interpolation stencil, the velocity on the Lagrangian points $\bm{X}_k$ is given by
\begin{equation}
    \bm{u}^*(\bm{X}_k, t) = \sum_{\bm{x}} \bm{u}^*(\bm{x}, t) W(\bm{x} - \bm{X}_k) \Delta x^3, 
\end{equation}
where the Dirac function
\begin{equation}
    W(x,y,z) = \frac{1}{\Delta x^3} w(\frac{x}{\Delta x})w(\frac{y}{\Delta x})w(\frac{z}{\Delta x}), 
\end{equation}
is the product of the 1D weight function
\begin{equation}
    w(r) = \left\{ 
    \begin{array}{ll}
        \frac{1}{8}\left(3-2|r|+\sqrt{1+4|r|-4r^2}\right) & \text{if } |r| \leq 1, \\
        \frac{1}{8}\left(5-2|r|-\sqrt{-7+12|r|-4r^2}\right) & \text{if } 1 \leq |r| \leq 2, \\
        0 & \text{otherwise} .
    \end{array} \right.
\end{equation}

The body force $\bm{f}_{imm}$ is determined by the following iterative procedure \citep{Ota2012LiftSimulations}:
\begin{enumerate}
    \item[\textbf{Step 0.}] Initial estimation of the body force on the Lagrangian points by
        \begin{equation}
            \bm{f}_0(\bm{X}_k, t) = \frac{\bm{U}_k(t) - \bm{u}^*(\bm{X}_k, t)}{\Delta t} .
        \end{equation}
    \item[\textbf{Step 1.}] Compute the body force on the Eulerian grid at the $l$th iteration by considering the nearby Lagrangian points
        \begin{equation}
            \bm{f}_l(\bm{x}, t) = \sum_{\bm{X}_k} \bm{f}_l(\bm{X}_k, t) W(\bm{x} - \bm{X}_k) A_k ,
        \end{equation}
    where $A_k$ is the corresponding surface area of the Lagrangian point $k$.
    \item[\textbf{Step 2.}] Correct the lattice velocity
        \begin{equation}
            \bm{u}_l(\bm{x}, t) = \bm{u}^*(\bm{x}, t) + \bm{f}_l(\bm{x}, t) \Delta t ,
        \end{equation}
    \item[\textbf{Step 3.}] interpolate the corrected velocity at the Lagrangian points
        \begin{equation}
            \bm{u}_l(\bm{X}_k, t) = \sum_{\bm{x}} \bm{u}_l(\bm{x}, t) W(\bm{x} - \bm{X}_k) \Delta x^3,
        \end{equation}
    \item[\textbf{Step 4.}] update the body force of Lagrangian point $\bm{X}_k$
        \begin{equation}
            \bm{f}_{l+1}(\bm{X}_k, t) = \bm{f}_l(\bm{X}_k, t) + \frac{\bm{U}_k(t) - \bm{u}_l(\bm{X}_k, t)}{\Delta t} ,
        \end{equation}
    \item[\textbf{Step 5.}] Repeat from \textbf{Step 1}.
\end{enumerate}
The above iterative procedure can be applied for as many cycles as the application demands. For the case of a single red blood cell, even one cycle gives satisfactory results. The body force computed by the last cycle on the Eulerian grid is the $\bm{f}_{imm}$ of algorithm \ref{algo:LBM}.

The immersed body force term $\bm{f}_{imm}$ could be used as the $\bm{F}_{ext}$ for the solid solver. However, we make use of the hydrodynamic stress tensor, equation \eqref{eq:stressTensor}, for higher accuracy. To compute the force at the Lagrangian point $k$, we project the stress tensor $\bm{\sigma}$ onto the surface normal $\bm{n}_k$ and we get
\begin{equation} \label{eq:Fext}
    \bm{F}_k^{ext} = \rho_{fluid} A_k \left ( \sum_{\bm{x}} \bm{\sigma}(\bm{x}) W(\bm{x} - \bm{X}_k) \right )  \bm{n}(\bm{X}_k).
\end{equation}

At this point we should highlight that in the immersed boundary method the same fluid covers the whole computational domain. This contradicts with the existence of interior fluid in the RBCs, of different viscosity and density, than the exterior blood plasma. However, for our study we disregard the contribution of the interior fluid on $\bm{F}_k^{ext}$ by summing only the points $\bm{x}$ that are on the outside region of the bodies. For increased stability, we found out that instead of summing the contribution of the exterior Eulerian points $\bm{x}$, it is sufficient to compute the force by considering the stress tensor of the most distant point from the Lagrangian point. Regarding the interior fluid, we follow the modeling approach described in section \ref{sec:npFEMRecap}, taking into account both the incompressibility and viscosity of cytoplasm, and thereby leaving the interior flow generated by the IBM as an artifact flow with no significance for our simulations.

{\color{black}
\section{Computational Framework} \label{sec:RECAP}
After the thorough presentation of each constituent of our framework, namely solid, fluid and FSI solvers, it would be appropriate to put all these pieces together and see the exact order of execution for one time step:
\begin{enumerate}
    \item Compute the macroscopic fluid properties: density ($\rho$), momentum ($\rho \bm{u}$) and stress tensor ($\bm{\sigma}$) from equations \eqref{eq:LBM_dens},\eqref{eq:LBM_Velocity},\eqref{eq:stressTensor}
    \item Using the stress tensor ($\bm{\sigma}$) of the previous step, compute the external forces on the solids from the fluid, i.e., $\bm{F}^{ext}(t)$ from equation \eqref{eq:Fext} \label{itm:Fext_cols}
    \item Apply the immersed boundary method (IBM) as presented in section \ref{sec:IBM}
    \item Impose the immersed boundary force ($\bm{f}_{imm}$) and any other forcing term to the fluid through the Shan-Chen forcing scheme as presented in algorithm \ref{algo:LBM}
    \item Collide \& Stream, lattice Boltzmann (LB) step as presented in section \ref{sec:LBM}, thus advance the fluid from $t$ to $t+1$
    \item Using $\bm{F}^{ext}(t)$ solve the dynamics of the immersed bodies (blood cells) as shown in algorithm \ref{algo:npFEM}, thus advance the solids from $t$ to $t+1$ \label{itm:npFEM_step}
\end{enumerate}
The steps above advance the whole system from $t$ to $t+1$. These steps are repeated for $m$ times such that $m \cdot \Delta t = $ physical simulated time needed. As stated by MacMeccan and colleagues \citep{Macmeccan2009SimulatingMethod}, the FE solver time step can be generally longer than the lattice time step. As a result, step \ref{itm:npFEM_step} can be called every $n_{npFEM}$ steps. However in the current work, we set $n_{npFEM} = 1$ to avoid complicating the validation of our framework. Furthermore, in the case of multiple bodies we have to consider collisions to avoid interpenetrations. The method is described in section \ref{sec:ProofOfCapability} and this step is realized together with step  \ref{itm:Fext_cols}.
}

\section{Results} \label{sec:Results}
In this section, we present a series of \emph{in silico} experiments and compare them with their \emph{in vitro} counterparts. A complete validation is thus performed, examining thoroughly all the constituents of the proposed framework. More in details, we validate the proposed RBC model disregarding any external fluid, by testing a RBC under stretching and recovery in sections \ref{sec:StretchingExp} and \ref{sec:RecoveryExp}. Following the validated deformable body model, we test the fluid solver and the \emph{fluid-structure interaction} (FSI), placing a single RBC in shear and Poiseuille flows in sections \ref{sec:TTExp} and \ref{sec:PoiseuilleExp}, while measuring suitable deformation indices. We then compare our framework against other state-of-the-art solvers in section \ref{sec:Benchmark}, and finally in section \ref{sec:ProofOfCapability} we demonstrate the capability of our framework to cope with multiple deformable bodies, which means, its ability to examine the collective transport of cells in the blood plasma for our future research endeavours.

The tuning of the deformable RBC model parameters presented in this section is minimal, since we are using well-known and tested materials to describe its mechanical and dynamical properties. Most importantly, the same set of parameters which is independent of the mesh resolution or regularity is used for all the \emph{in silico} experiments.

{\color{black}
Our framework is written in C\texttt{++}, it utilizes Palabos \citep{PalabosHttp://www.palabos.org} for the IB-LBM parts and our own npFEM library for the solid part. The npFEM library is based on ShapeOp \citep{ShapeOpWebSite} which is an implementation of projective dynamics, freely available on the web. Currently, the complete framework is written for CPUs with extensive use of Message Passing Interface (MPI), which deals with the distribution of the load to all the available cores. In forthcoming work, we will present a hybrid framework where the IB-LB parts are executed on CPUs while the npFEM solver is fully deployed on the available general purpose Graphics Processing Units (gpGPUs). Extensive performance tests are to be expected with this work.

For an easier overview, we provide table \ref{tab:Params} that consolidates the most critical parameters used in our numerical experiments.
}

\begin{table}
\color{black}
\centering
\caption{Overview of used parameters}
\begin{tabular}{lc}
\hline
\textbf{Parameter} & \textbf{Value} \\
\hline
$\Delta x$ & $0.5~\mu m$ \\
relaxation time $\tau$ & $[1,2]$ \\
$\Delta t$ & equation \eqref{eq:diffusiveScaling} \\
Skalak B,C,D & $5, 5000, 35~pN/\mu m$ \\
$k_{bending}$ & $1~pN \mu m$\\
$k_{volume}$ & $250$ \\
Rayleigh $\alpha_D, \beta_D$ & $0, 0.01$\\
$\kappa_{\text{damping}}$ & $0.6$ \\
IBM cycles & $[1,5]$ \\
RBC mesh resolution (surface vertices) & $\{ 66, 130, 258, 514, 719, 1026 \}$ \\
\end{tabular}
\label{tab:Params}
\end{table}

\subsection{Stretching experiment} \label{sec:StretchingExp}
The \emph{in vitro} stretching of RBCs is performed using the optical tweezers technique (Dao et al. 2003/ 2006 \citep{Dao2003MechanicsTweezers, Dao2006MolecularlyErythrocyte}, Mills et al. 2004 \citep{Mills2004} and Suresh et al. 2005 \citep{Suresh2005ConnectionsMalaria}), where small silica beads are attached on opposite sides of the membrane and are moved by a laser beam that traps and manipulates them. The resulting deformation leads to an extension along the longitudinal axis and compression along the transverse axis. To simulate the above stress and strain condition, we apply equal and opposite forces on selected vertices of the RBC membrane. The force is increased gradually up to $200~pN$ (pN: pico Newton). Force increments are applied only when the red blood cell reaches an equilibrium state. The viscous properties of the membrane cannot be tested in this experiment, since we are interested in the converged equilibrium state only. The measurements taken to assess the deformation in this kind of experiment are the axial and transverse diameters.

\begin{figure}[h]
    \centering
    \includegraphics[scale=0.5]{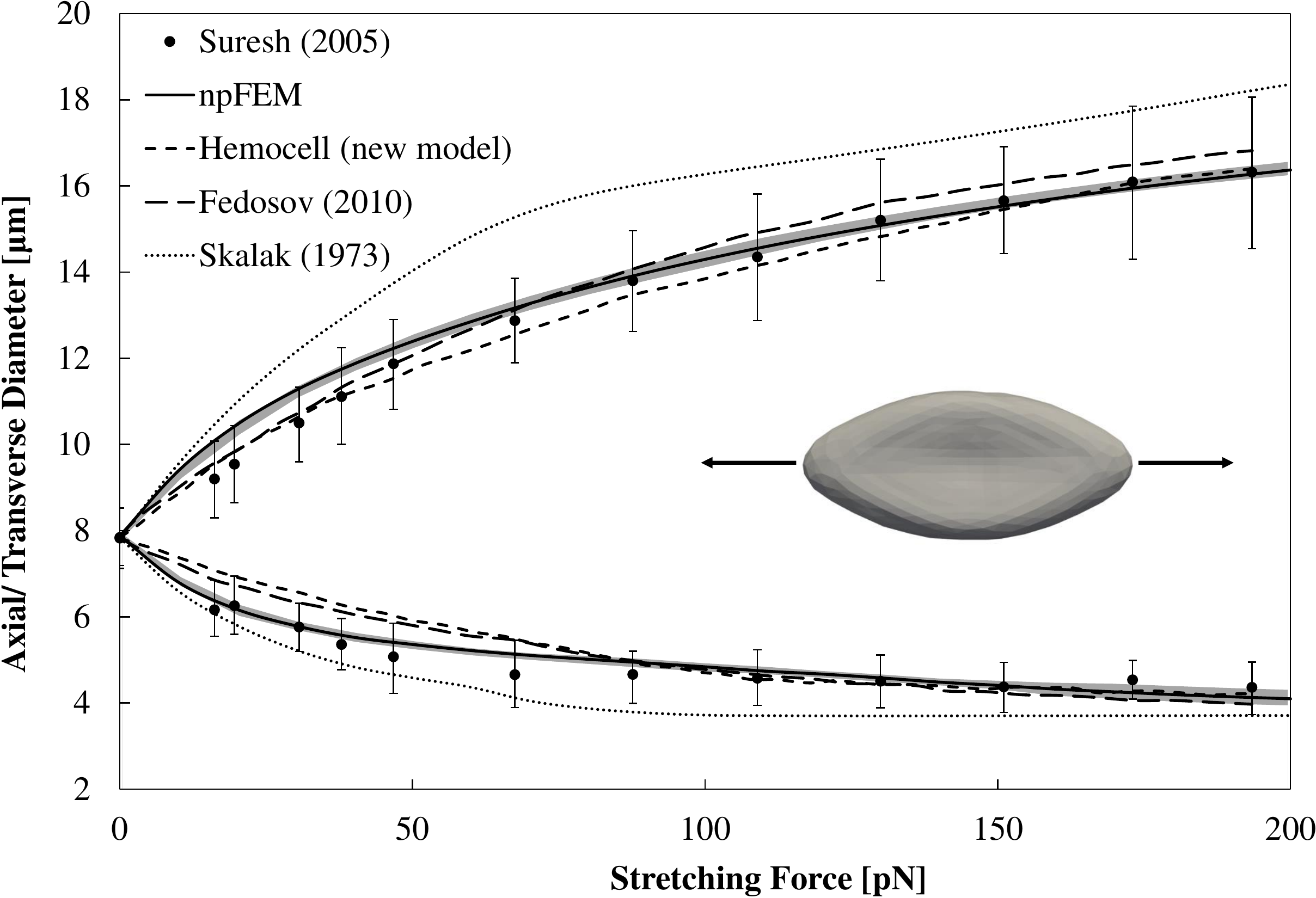}
    \caption{Stretching Experiment: Comparison of our solver (npFEM) with the experiments of Suresh et al. \citep{Suresh2005ConnectionsMalaria} and other numerical solvers. The correction of Skalak's model \citep{Skalak1973StrainMembranes} leads to superior strain-hardening behavior compared to the original. The sub-figure presents a stretched RBC as produced by the npFEM solver.}
    \label{fig:Stretching_Graph}
\end{figure}

The simulation of the RBC includes all the energy terms presented in section \ref{sec:npFEM}, namely the ones representing area and volume conservation, as well as the bending and the material energies. There are not many free parameters to be tuned, since we adopt the values from the original model proposed by Skalak \citep{Skalak1973StrainMembranes}, which are $B=0.5~10^{-2}~dyn/cm = 5~pN/\mu m$ and $C = 5~dyn/cm = 5000~pN/\mu m$. The $C$ parameter corresponds to the stiffness $k_i$ of the weight $w_i (= A_i k_i)$ of the area preservation energy presented in \ref{sec:QuadraticEnergies}. For the bending rigidity of the membrane we use an augmented value from the one presented in \citep{Fedosov2010ARheologydynamics,Zavodszky2017CellularCells}, i.e., $k_{bending} = 2 \widetilde{\kappa} = 4 \kappa / \sqrt{3}  \approx 1~pN \mu m$, where $\kappa = 100 \kappa_B T$. The weight for the volume preservation can take any arbitrary, sufficiently high value. In our simulations, we chose $k_{volume} = 250$. The only free parameter to be tuned is $D$, which is the stiffness of the correction term in Skalak's energy (see section \ref{sec:correctedSkalak}). We found out that $D = 35~pN/ \mu m$ results in a close fitting to the experimental results of \citep{Suresh2005ConnectionsMalaria}. As shown in Figure \ref{fig:Stretching_Graph}, our model (npFEM - continuous line) presents a fit of better quality than other solvers, especially concerning the transverse diameter at low deformations, where most of the solvers fail. The modification of Skalak's model adds the energy term $(\lambda_i - 1)^4 D/4$ in the energy density function $\Psi$. This term is mainly activated whenever the stretching ratios $\lambda_i$ diverge from the rest state ($\lambda_i = 1$). Therefore, it improves the strain-hardening behavior, whereas the original model fails to fit the experimental data. The set of parameters defined here is used for any other case study presented in the results section and does not depend on the underlying mesh structure.

We represent our experimental results by a shaded region (Figure \ref{fig:Stretching_Graph}), which encompasses the output from all the tested cases with varying RBC mesh resolution, i.e., from 66 surface vertices up to 1026. The continuous line within the shaded region represents the case with $719$ surface vertices. It is obvious that our solver is characterized by strong \emph{mesh independence} which is a direct consequence of the proper discretization of the continuous energies ({\color{black} a general advantage of FEM-based solvers}). Currently, all the state-of-the-art solvers \citep{Fedosov2010ARheologydynamics, Zavodszky2017CellularCells} are based on coarse-grained spectrin-link models which are essentially mass-viscoelastic spring systems. Their calibration is based on a method first introduced by Dao et al. 2006 \citep{Dao2006MolecularlyErythrocyte} and later extended by Fedosov et al. 2010 \citep{Fedosov2010ARheologydynamics}, which relies on a regular two-dimensional sheet of springs, i.e., equilateral triangular elements. In this case, the physical parameters are linked to the mesh resolution. A disadvantage is that coarser meshes cannot maintain mesh regularity and therefore these methods fail to maintain mesh independence, demanding tedious tuning of the parameters set.

{\color{black}
The time discretization is irrelevant in this kind of experiments since we are interested on the converged state of the stretched RBC. Using the backward Euler integration scheme, we guarantee unconditional stability of the simulation under any unfavorable situation, e.g., arbitrary large time step, large deformations, extreme mesh underresolution. Indeed, Figure \ref{fig:Stretching_Graph} validates this statement and shows not only how stable is the npFEM solver (under-resolved case of 66 surface vertices), but also its accuracy since all the different simulations fall very close to the reference solution (experimental data). 
}

{\color{black}
As stated by Siguenza and colleagues \citep{Siguenza2017HowMechanics}, the axial and transverse diameters measured in optical tweezers experiments are found to be insufficient for validation purposes. They propose other quantities such as the height or the profile of the cell, which seem to be more sensitive to different membrane models. Nevertheless, they conclude and agree with the findings of Dimitrakopoulos et al. 2012 \citep{Dimitrakopoulos2012AnalysisModeling} that the only constitutive law able to properly match the wide variety of experimental data available in the literature is Skalak's constitutive law, a variation of which we are using for our simulations. Additionally, we decided not to include other measured quantities as it is very difficult to find these deformation indices in the literature and compare with.
}

\subsection{Recovery experiment} \label{sec:RecoveryExp}
Stretching experiments help us define some of the most important mechanical properties of the RBC, but they do not reveal information about the viscoelastic nature of the membrane. There are two experiments that contribute towards this direction, namely the recovery and the tank-treading. The initial stage of the recovery experiment is the axial deformation of the RBC by optical tweezers \citep{Henon1999AErythrocyte}, just like in the stretching experiment. When the deformation reaches the desired value, the forces are released and the time until the RBC reaches the initial state (rest shape) is tracked. To study the recovery experiment, Hochmuth et al. 1979 \citep{Hochmuth1979RedViscosity} introduced the time dependent elongation index
\begin{equation} \label{eq:ElongationIndex}
    e(t) = \frac{(\lambda - \lambda_{\infty}) (\lambda_0 + \lambda_{\infty})}{(\lambda + \lambda_{\infty}) (\lambda_0 - \lambda_{\infty})} = exp \left [ - \left ( \frac{t}{t_c} \right )^{\delta} \right ],
\end{equation}
where $\lambda = D_A / D_T$ ($D_A$ and $D_T$ are the axial and transverse diameters), $\lambda_0$ and $\lambda_{\infty}$ correspond to the ratios at time $t=0$ and $t=\infty$, $t_c$ is the characteristic recovery time, and $\delta$ is the exponent as suggested by \citep{Fedosov2010ARheologydynamics}. As shown by Hochmuth et al. 1979 \citep{Hochmuth1979RedViscosity} the time rate of recovery is related mainly to the membrane viscosity and much less to the RBC cytoplasm or the outside aqueous environment. Given this observation, the numerical experiments are performed disregarding the existence of surrounding fluid. 

\begin{figure}[h]
    \centering
    \includegraphics[scale=0.5]{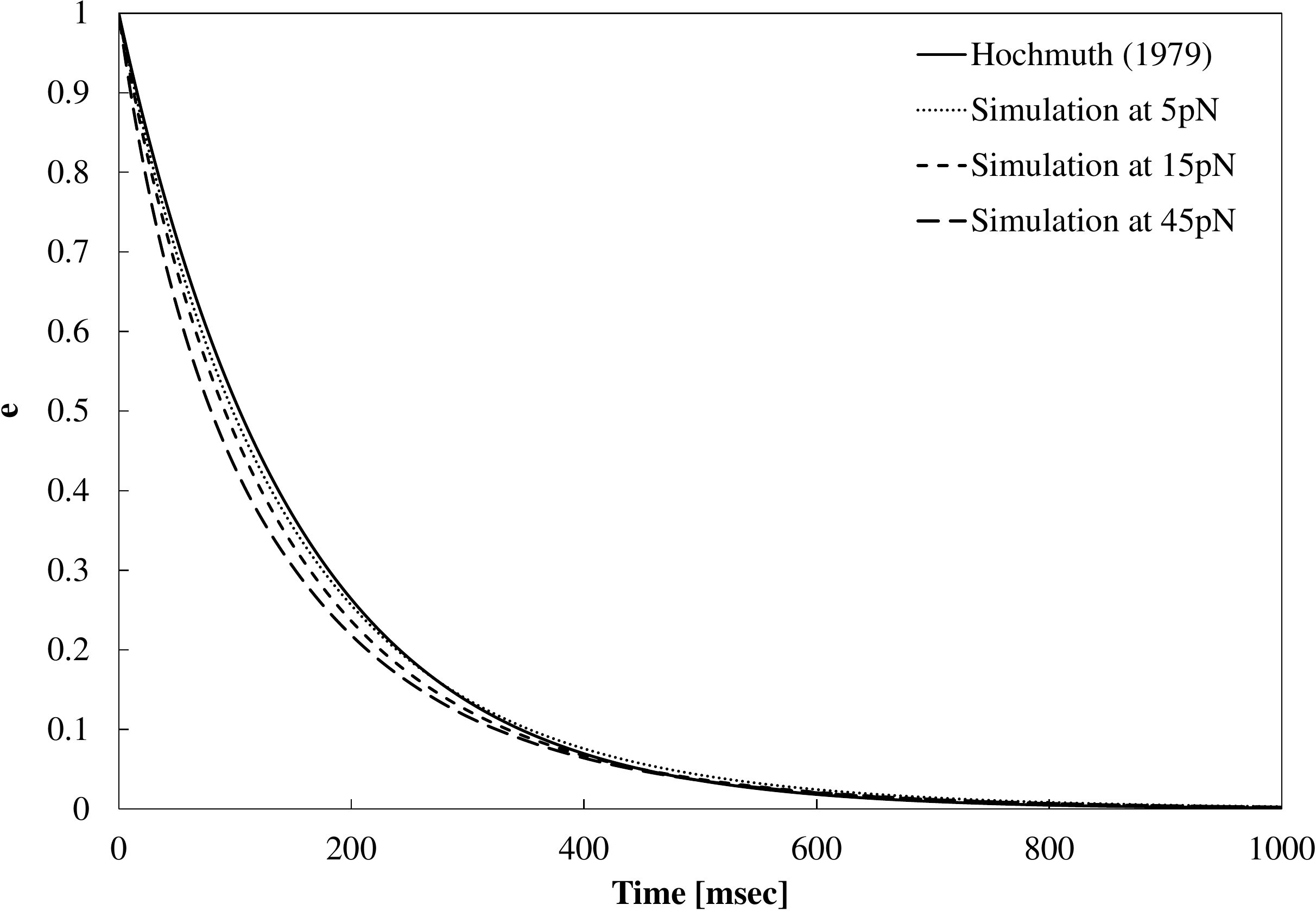}
    \caption{RBC recovery after deformation at various stress levels and comparison with the theoretical curve (for $\delta = 0$ and $t_c = 150~msec$). The divergence from the theoretical curve at higher stress levels is due to non-linear effects.}
    \label{fig:Recovery_Graph}
\end{figure}

As observed from Figure \ref{fig:Recovery_Graph}, the emulated viscoelastic model described in section \ref{sec:Viscoelasticity} fits the curve of equation \eqref{eq:ElongationIndex} very accurately, for $\delta = 0$ and $t_c = 150~msec$. As the stress levels increase, non-linear effects grow and a deviation from the theory is unavoidable, since it was developed for smaller deformation levels. Hochmuth suggested $t_c \approx 100 - 130~msec$, while H\'enon et al. 1999 \citep{Henon1999AErythrocyte} suggested a characteristic time in the range of $100 - 300~msec$ where the lower bound corresponds to younger RBCs and the upper to older. Our tuning describes a RBC somewhere in the middle with $t_c = 150~msec$, $\beta_D = 0.01$ and $\kappa_{\text{damping}} = 0.6$.

In the mass-spring solvers \citep{Fedosov2010ARheologydynamics, Zavodszky2017CellularCells}, the viscosity of the RBC membrane is tackled through the addition of a dissipative force $\bm{F}_{ij}^D = - \eta \bm{v}_{ij}$, where $\eta$ is the membrane viscosity and $\bm{v}_{ij}$ is the relative velocity of the spring ends. However, this forcing term has the same disadvantage as the Rayleigh damping, i.e., it damps rigid body motions instead of only the relative ones. For example, a pure rotation produces $\bm{v}_{ij} \neq 0$, and thus internal forces are produced, even if they should not. This inherent drawback of $\bm{F}_{ij}^D$ has a severe impact on the dynamics of RBCs. A possible remedy would be to project the velocity difference $\bm{v}_{ij}$ onto the vector separating the masses. 

\subsection{Wheeler experiment} \label{sec:WheelerExp}
In this experiment, as performed by Yao et al. 2001 \citep{Yao2001LowChamber}, RBCs are positioned in a shear flow with their axis of symmetry perpendicular to the flow velocity. The experiment, also known as wheeler experiment, takes its name from the characteristic wheel shape of the deformed RBCs. The red blood cells are submerged in a phosphate buffered solution (PBS) of low dynamic viscosity ($0.707 cP$) and are submitted to a shear flow with shear rates in the range of $15 - 200~s^{-1}$. The deformation of the RBCs is quantified via the deformation index (DI) given by
\begin{equation}
    DI = \frac{(D_{max}/D_0)^2 - 1}{(D_{max}/D_0)^2 + 1},
\end{equation}
where $D_0$ is the initial RBC diameter ($\sim 7.82~\mu m$) and $D_{max}$ is the maximal diameter after deformation at constant shear rate. This is the first from a series of experiments that employs the fluid-structure interaction framework, which is composed of the lattice Boltzmann (LB) fluid solver, the nodal projective Finite Element (npFE) solid solver and the immersed boundary (IB) method for the coupling of the two solvers.

\begin{figure}[h]
    \centering
    \includegraphics[scale=0.5]{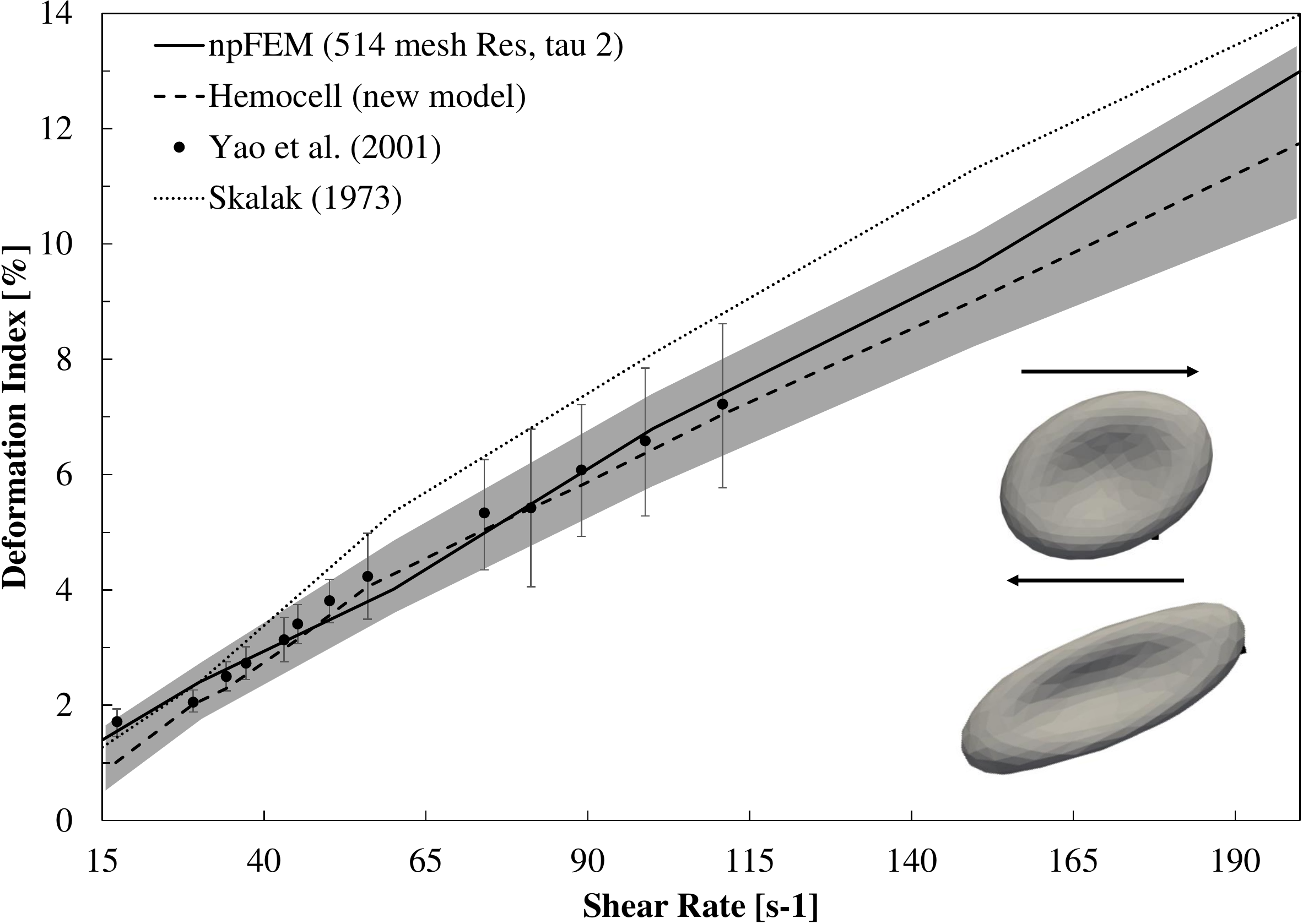}
    \caption{Wheeler experiment: Comparison of our framework with the experimental data of Yao et al. 2001 \citep{Yao2001LowChamber} and Hemocell solver \citep{Zavodszky2017CellularCells}. The sub-figures show deformed RBCs under $200~s^{-1}$ (upper) and $2000~s^{-1}$ (lower) shear rates as produced by our computational framework.}
    \label{fig:Wheeler_Graph}
\end{figure}

The RBC is placed at the center of the tank ($20\times20\times20~\mu m^3$) with periodic boundary conditions except for the walls. The wall velocity tunes the shear rate to the desired one. The spatial discretization is $\Delta x=0.5~\mu m$, the relaxation time $\tau \in [1,2]$, and the time step is defined through the diffusive scaling formula as
\begin{equation} \label{eq:diffusiveScaling}
    \Delta t = c_s^2 \left ( \tau - \frac{1}{2} \right ) \frac{\Delta x}{\nu},
\end{equation}
where $\nu$ is the kinematic viscosity of the ambient fluid. {\color{black} A remark at this point is that in lattice Boltzmann simulations, if one fixes the spatial discretization and the relaxation time for a given fluid (known viscosity), then there is no freedom over the selection of the time discretization, as it is dictated by the diffusive scaling formula. If $\Delta t$ is not defined through equation \eqref{eq:diffusiveScaling}, then physics are violated. However, in cases where the particle Reynolds number and the capillary number are very small, then a violation of $\Delta t$ can lead in an acceptable deviation of the simulated dimensionless numbers from the real ones. For a detailed overview on selecting model parameters for LBM simulations, we suggest chapter 8 from Kr{\"{u}}ger 2012 \citep{KrugerThesis}.}

As seen from Figure \ref{fig:Wheeler_Graph}, our framework (continuous line) fits very accurately the experimental data \citep{Yao2001LowChamber} {\color{black}(reference solution to investigate accuracy of the solver)}. The continuous line corresponds to a simulation of $514$ surface vertices, $\tau = 2$ and $l=5$ iterations for the IBM algorithm. The shaded region around our results encompasses all the different case studies that we simulate with varying RBC mesh resolution (from 66 to 1026 surface vertices), different values of $\tau \in [1,2]$ and different iteration steps for the IBM iterative algorithm $l \in [1,5]$ as presented is section \ref{sec:IBM}. {\color{black} This region is wider than the one of Figure \ref{fig:Stretching_Graph} where we do not have fluid-solid interaction, nonetheless the property of mesh independence is validated once more, since all the different experiments fall very close to the reference solution. The accuracy of the simulations seems more sensitive when the fully coupled framework is deployed. Essentially, the immersed boundary method introduces stricter requirements on the resolution of the different solvers and how the resolutions match with each other. Nevertheless, for this single-cell case study even the under-resolved case with 66 surface vertices seems to be very close to the reference solution by Yao. At this point, we have to mention the work by Kr{\"{u}}ger 2012 \citep{KrugerThesis} where the author suggests that the ratio $\bar{l}/\Delta x$, where $\bar{l}$ is the average edge length of the membrane mesh, should be somewhere between $0.5$ and $1.5$. The RBC membrane with $258$ vertices satisfies the upper bound of the above requirement and clearly the cases with $66$ or $130$ vertices violate this criterion. In spite of this violation, the framework accurately reproduces the reference solution. On the other hand, for cases of higher hematocrit ($>30\%$) we expect to face accuracy issues and probable instabilities. More in details, IBM demands $\bar{l}/\Delta x \sim 1$ to avoid fluid penetrating the solid surfaces and also mesh under-resolution leads to poorly resolved collisions between bodies. This topic will be thoroughly investigated in upcoming work.}

The proposed solver presents also great robustness, since it withstands shear rates as high as $2000~s^{-1}$ without any stability issues (see Figure \ref{fig:Wheeler_Graph}). Once more, Skalak's original formulation presents a correct behavior for small deformations but fails for higher shear rates. The modification term, which is significant for larger deformations, repairs the RBC behavior within the whole deformation range. It is important to point out that a modification like the one presented by our framework cannot be easily realized by the coarse-grained spectrin link solvers, because the spring parameters are not intuitively linked to the material properties. On the contrary, the npFEM solver allows great material expressivity with minimum effort from the side of the user.

\subsection{Tank-Treading experiment} \label{sec:TTExp}
Experimental observations of RBCs in shear flows \citep{Fischer894, Fischer2007Tank-TreadMedium, Tran-Son-Tay1984DeterminationMotion, Abkarian2007SwingingFlow} reveal a tumbling (T) motion at low shear rates and tank-treading (TT) at higher shear rates. In this experiment, the RBC's axis of symmetry is oriented along the flow velocity and its behavior is examined for different shear rates in the tank-treading regime. The ambient fluid dynamic viscosity ($5~cP$) is much larger than in the wheeler experiment. We are following the same experimental setup as Fedosov et al. 2010 \citep{Fedosov2010ARheologydynamics} and Tran-Son-Tay et al. 1984 \citep{Tran-Son-Tay1984DeterminationMotion}. The RBC (258 vertices mesh resolution) is positioned at the center of a domain $20 \times 20 \times 20~\mu m^3$, the spatial resolution is $0.5~\mu m$ and the relaxation time $\tau=1$. The IBM iterations can vary from one to five without any difference on the resulting deformation.

\begin{figure}[h]
    \centering
    \includegraphics[scale=0.5]{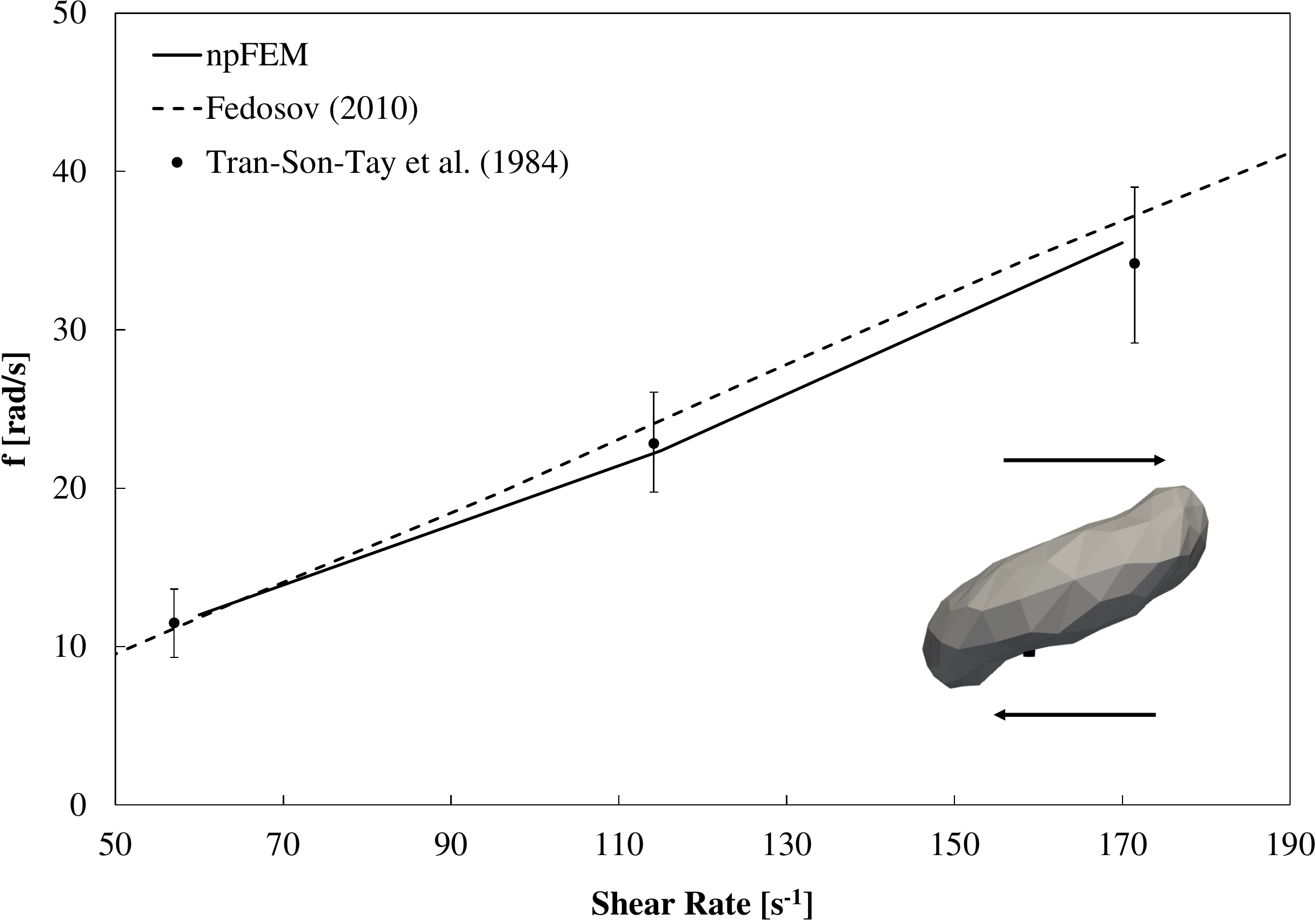}
    \caption{Tank-treading frequency of a RBC in shear flow. Experimental setup as described in \citep{Tran-Son-Tay1984DeterminationMotion}. The sub-figure presents a RBC exhibiting tank-treading as produced by our framework (RBC mesh resolution: 258 vertices). Note that the investigated regime focuses on tank-treading rather than tumbling.}
    \label{fig:NoneWheelerGraph}
\end{figure}

\begin{figure}[h]
    \centering
    \includegraphics[width=\textwidth]{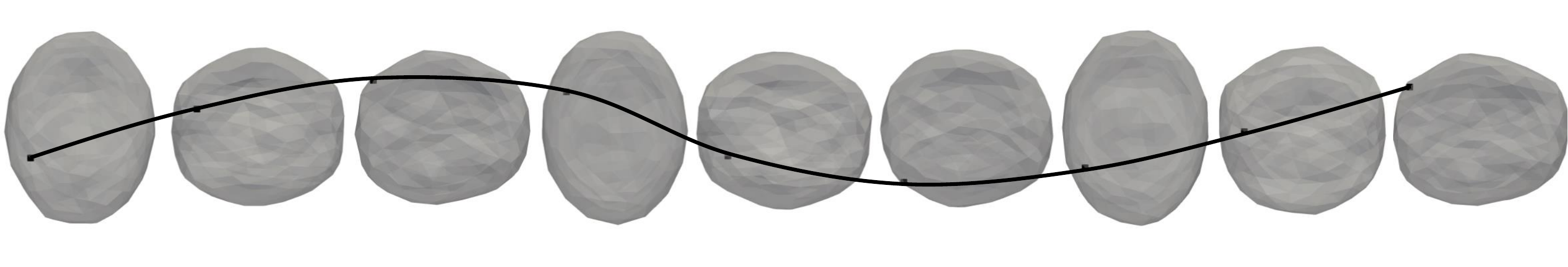}
    \caption{Simulated tank-treading (TT) by our framework. We gave a transparency to the RBC in order to track the marker whenever it is located at non-visible places.}
    \label{fig:TankTreading}
\end{figure}

Tran-Son-Tay et al. 1984 \citep{Tran-Son-Tay1984DeterminationMotion} studied this setup and defined two curves of the TT frequency (TTF) for young and old RBCs. In the present study, we are averaging the two curves and the ``error'' bounds correspond to the maximum and minimum values. The TTF is calculated by tracking a bead attached on the RBC surface and counting the cycles per second (see Figure \ref{fig:TankTreading}). Figure \ref{fig:NoneWheelerGraph} shows an excellent agreement of our framework compared to the experimental data of \citep{Tran-Son-Tay1984DeterminationMotion}. Figure \ref{fig:TankTreading} shows that the tank treading motion is very precisely captured, while the continuous curve is qualitatively the same as compared to Fischer et al. 1978 \citep{Fischer894}. The TT motion at higher shear rates results in a ``breathing'' of the RBC, i.e., in an alternating compression-extension in the longitudinal axis as shown in Figure \ref{fig:TankTreading}.

\subsection{Poiseuille flow experiment} \label{sec:PoiseuilleExp}
\begin{figure}[h]
    \centering
    \includegraphics[scale=0.5]{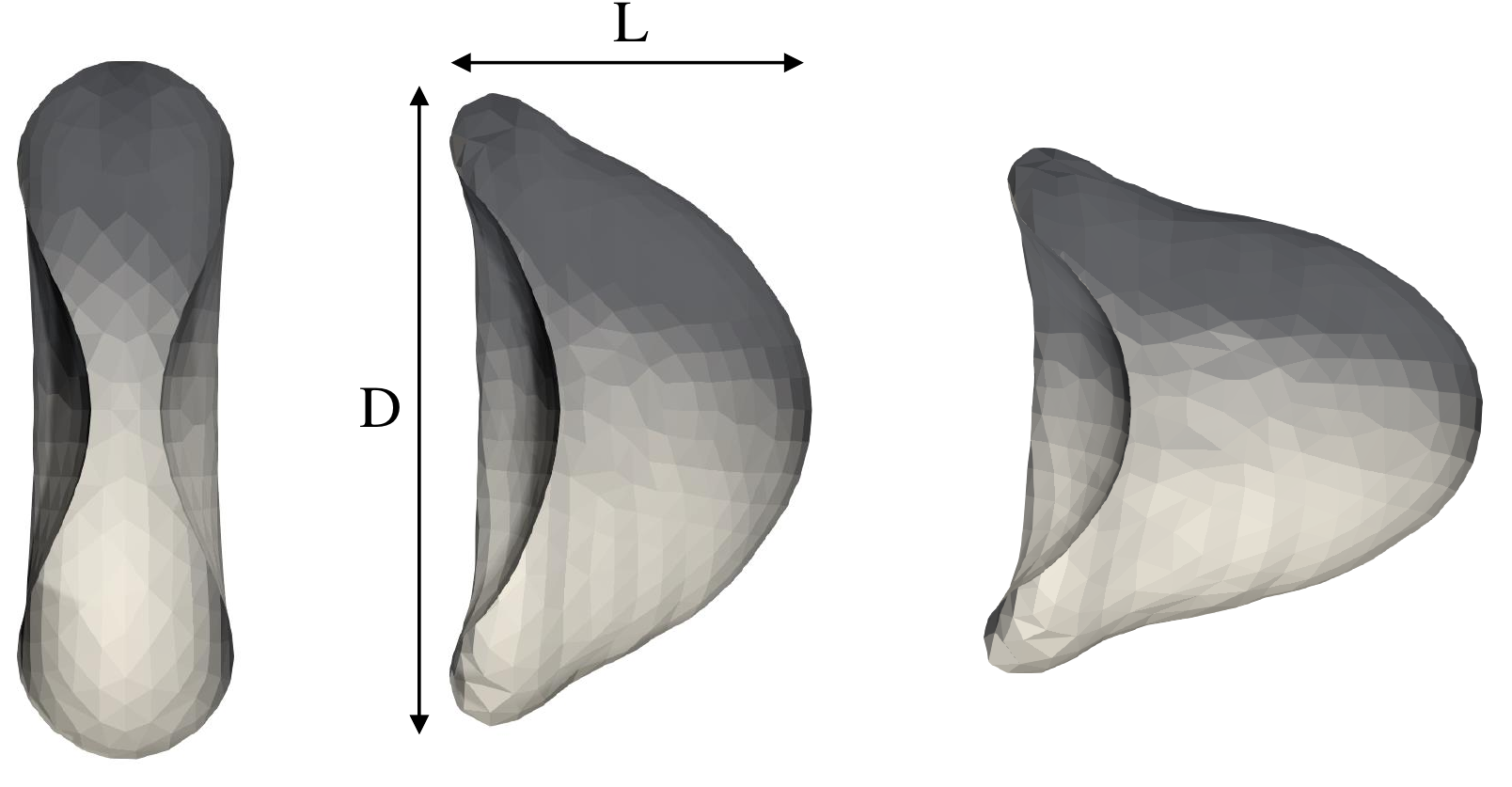}
    \caption{RBC transitioning to parachute-like shape in a tube with diameter $9.2~\mu m$ as produced by our framework.}
    \label{fig:Poiseuille}
\end{figure}

The ability of RBCs to travel through microcapillaries much smaller than their size is paramount for the transfer of oxygen to the organs. Thereby, the ability of a solver to simulate confined flows is a great advantage. Several researchers did experimental studies on the deformation of RBCs in small tubes and channels, e.g., Tomaiuolo et al. 2011 \citep{Tomaiuolo2011MicrofluidicsViscoelasticity}, Tsuakada et al. 2001 \citep{Tsukada2001DirectSystem}, and Suzuki et al. 1996 \citep{Suzuki1996DeformationDeformability}. The main feature of this flow is the transition of the RBC biconcave shape to a parachute-like shape, which intensifies as the velocity increases. A common way to quantify the deformation of the RBC in the parachute-like regime is via a deformation index defined as $\Gamma = L/D$, where $L$ is the length of the parachute (length of the dome) and $D$ is the deformed diameter, with an approximate initial value of $7.82~\mu m$. Tsukada et al. 2001 \citep{Tsukada2001DirectSystem} provide a graph where they correlate $\Gamma$ with the RBC velocity. Generally for tube diameters close to the RBC diameter, $\Gamma$ adopts values larger than 1. In our numerical setup, the tube diameter is $9.2~\mu m$, the RBC velocity varies from $4~\text{to}~9 ~mm/s$, and for all cases we achieved $\Gamma > 1$. In Figure \ref{fig:Poiseuille}, we present the transition from the biconcave to the parachute-like shape. The resulted shape is very close to the experimentally observed one by Tomaiuolo et al. 2011 \citep{Tomaiuolo2011MicrofluidicsViscoelasticity}. Simulations in confined spaces demand finer resolutions and smaller time steps to maintain stability, and our spatial discretization consquently is chosen to be $\Delta x = 1/3~\mu m$ with a relaxation time $\tau = 1$. These stability issues are not to be confused with the unconditional stability of the npFEM solver, since they refer to the FSI coupling. The \emph{explicit} coupling of the fluid solver with the solid npFEM solver introduces some limitations at extreme flow cases like the investigated one. With \emph{explicit} coupling, we mean that the external forces on the RBC come from the positions at time $t$ and not at $t+1$, see equation \eqref{eq:RayleighNewton} for more details.

The parachute-like shape is a challenging deformation to capture. Most of the solvers \citep{Zavodszky2017CellularCells, Fedosov2010ARheologydynamics} present a parachute with $\Gamma \ll 1$, where the dome is very small. Our solver on the contrary captures the expected shape even if the employed bending rigidity is higher than the suggested one.

{\color{black}
\subsection{Computational efficiency} \label{sec:Benchmark}
Main focus of our work is the solid solver, and thus the computational efficiency concerns the npFEM solver with respect to other well-known implementations. Currently, the state-of-the-art \emph{open-source} solver for the simulation of the collective transport of cells in blood plasma is the software Hemocell \citep{Zavodszky2017CellularCells, Zavodszky2017Hemocell:Library}. The solver uses Palabos \citep{PalabosHttp://www.palabos.org} for the blood plasma, the RBCs are represented as discrete element membranes (mass-viscoelastic spring systems based on the model introduced by \citep{Fedosov2010ARheologydynamics}), and the FSI is realized by a modified IBM introduced in Mountrakis 2015 \citep{Mountrakis2015TransportModels}. Their framework has a proven capability of simulating multiple deformable bodies, thanks in part to the computationally efficient solid structure solver. Moreover, other state-of-the-art high-performing solvers, e.g., Blumers et al. 2017 \citep{BLUMERS2017171}, Rossinelli et al. 2015 \citep{Rossinelli:2015:ILP:2807591.2807677}, use almost the same methods for the solid solver, i.e., coarse grained spectrin-link models, and thus we focus on Hemocell which is available to the public and has many similarities on the fluid \& FSI parts with our computational framework. Additionally, we avoid comparisons with GPU-accelerated solid solvers (see Blumers et al. 2017 \citep{BLUMERS2017171}) since it would not be fair to compare a CPU with a GPU implementation. In an upcoming project, we will present a GPU version of the npFEM solver that achieves execution time per iteration at the order of nanoseconds in contrast to the current efficiency which is at the order of milliseconds.

The reference simulation is chosen to be the wheeler experiment, presented in section \ref{sec:WheelerExp} and both solvers are executed on the same workstation. As expected, comparing an explicit mass-viscoelastic spring system with an implicit FEM-based solver, the efficiency comparison favors the former, as seen in table \ref{tab:BenchmarkWithHemocell}. Nevertheless, and in spite of the higher complexity of the approach, the npFEM framework appears to exhibit a computational efficiency quite close to Hemocell and consequently to any other mass-spring system solver.

It is also worth noting the benchmark of the linear explicit FE solver of MacMeccan et al. 2009 \citep{Macmeccan2009SimulatingMethod} for which the average computational time for an update is $2.3~msec/step$ for a RBC with $254$ surface vertices. A follow-up work from the same group, Reasor et al. 2011 \citep{Reasor2011CouplingFlow}, stated that the aforementioned solver was abandoned because of accuracy issues, since it could not capture tank-treading and the parachute-like shape in confined flows.

\begin{table}[h]
\centering
\begin{tabular}{|c|c|c|c|}
\hline
\textbf{SV: Surface Vertices}                      & Hemocell SV:642 & npFEM SV:258 & npFEM SV:667 \\ \hline
Average execution time per time step [ms/step] & 0.58            & 1.35         & 5.19         \\ \hline
\end{tabular}
\caption{Comparison of npFEM with Hemocell \citep{Zavodszky2017CellularCells, Zavodszky2017Hemocell:Library}. The timing refers to the deformable bodies solvers.}
\label{tab:BenchmarkWithHemocell}
\end{table}
}

\subsection{Proof of Capability} \label{sec:ProofOfCapability}
Recapitulating the performance tests of our solver, one can observe that the extension to flows of multiple RBCs should not be a problem. Indeed, with a computational time per update very close to the one of mass-spring systems, and with proven capability of the latter to scale well with increasing number of deformable bodies \citep{Mountrakis2015ParallelFramework}, we present Poiseuille and shear flows with O($10^2$) RBCs. The deliberately small number of RBCs helps us present the rich dynamics of individual RBCs produced by our framework. Studies with larger numbers of RBCs and higher hematocrit will be part of upcoming work. {\color{black} Furthermore, dealing with vessels of few micrometers, the hematocrit is by definition low. As stated by Zhao \& Shaqfeh 2011 \citep{zhao_shaqfeh}, although the normal range of human body hematocrit is 40\%-50\%, the local hematocrit drops significantly in small vessels and is only about 40\% of the body average at $30~\mu m$ vessel diameter. Hence the mean hematocrit in this study is chosen to be around 20\%.}

To run a case with multiple RBCs in a robust manner, we need on top of our framework a collision detection/ handling system. Regarding the collision detection, there are two stages. In the first, Palabos \citep{PalabosHttp://www.palabos.org} uses an efficient data structure for identifying, for each vertex of the examined RBC, neighboring vertices on other RBCs. The algorithm is based on the idea that all RBC vertices are attached to a closely located fluid cell, allowing to efficiently locate them on ground of their spatial position. At the second stage, npFEM solver builds a local kd-tree \citep{blanco2014nanoflann} for an even faster collision detection system, reducing the colliding data received from Palabos. The collision handling fits naturally in our solver in the sense that whenever a collision is detected, a collision energy/ force is added to the non-PD part of our solver (see equations \eqref{eq:GeneralizedObjectiveFunction} and \eqref{eq:GeneralizedObjectiveFunctionDerivative}) acting upon the vertex under collision. Every colliding point has a desired position ($\bm{p}_i$) which is defined from the colliding neighbors and their normals. Thus the collision energy and force are defined as
\begin{align}
    E_i^{non-PD} &= \frac{w_{col}}{2}|| \bm{p}_i - \bm{x}_i ||_F^2, \\
    F_i^{non-PD} &= w_{col} (\bm{p}_i - \bm{x}_i),
\end{align}
where $w_{col}$ is the weight of the collision term in the optimization problem. An important precaution is that at every iteration of the solver, the algorithm checks if $\bm{x}_i$ still penetrates the neighboring body. If it does not, the corresponding collision term is cancelled. This procedure avoids sticking of the bodies, which would be unrealistic. The collision energies and forces have no contribution to the Hessian approximation matrix ($\widetilde{\mathbf{M}}/h^2 + \mathbf{L}$) and thus they may increase the convergence steps. However, after extensive testing, we found out that because of the small time steps (limitation introduced by the explicit nature of LBM), the collision handling process has no measurable impact on the total computational time. While the suggested collision energies favor simplicity over accuracy, any other type of energy can replace our simple spring-like equation. The efficient handling of collisions is one of the striking advantages of our method, as other frameworks such as Hemocell \citep{Zavodszky2017CellularCells} skip collision detection, and rely on a sufficient fluid resolution to avoid collisions thanks to lubrication effects.

Both Poiseuille and shear flows are executed on a regular lattice of size $32 \times 32 \times 80~\mu m^3$ with periodic boundary conditions except from the walls. The spatial discretization for the blood plasma is $0.5~\mu m$, the LBM relaxation time $\tau = 1$, the dynamic viscosity of the blood plasma $\sim 1.2~cP$, and the time discretization is computed from equation \eqref{eq:diffusiveScaling}.

\begin{figure}[h]
    \centering
    \includegraphics[width=\textwidth]{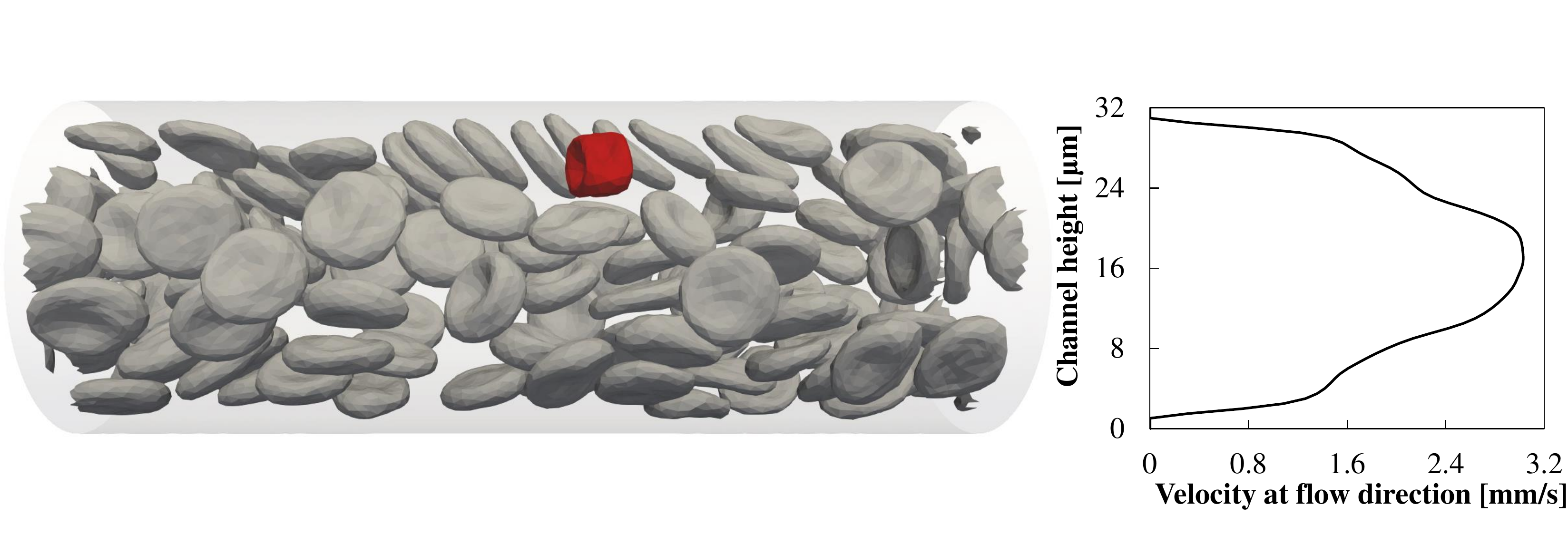}
    \caption{Simulated Poiseuille flow with no interpenetrations. The marked red blood cell (in red) has developed a multilobe shape as described in Lanotte et al. 2016 \citep{Lanotte13289}. The graph shows a temporal average of the velocity at the flow direction in the middle of the tube.}
    \label{fig:Poiseuille_MultipleRBCs}
\end{figure}

\begin{figure}[h]
    \centering
    \includegraphics[width=\textwidth]{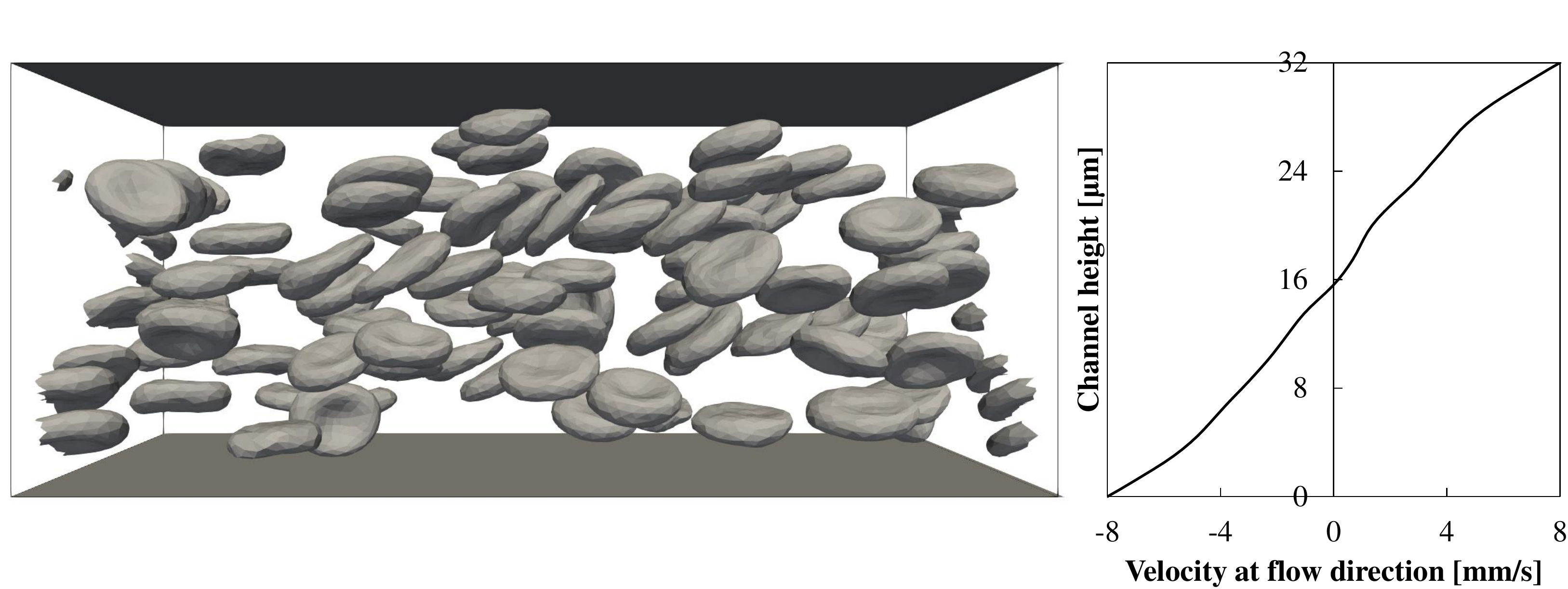}
    \caption{Simulated shear flow at shear rate $500~s^{-1}$. The graph shows a temporal average of the velocity at the flow direction in the middle of the channel.}
    \label{fig:Shear_MultipleRBCs}
\end{figure}

The Poiseuille flow as seen in Figure \ref{fig:Poiseuille_MultipleRBCs} consists of 123 RBCs (258 surface vertices each) inside a tube of $15.75~\mu m$ radius at a hematocrit of $\sim 20\%$. The shear flow as seen in Figure \ref{fig:Shear_MultipleRBCs} consists of 123 RBCs (258 surface vertices each) at a hematocrit of $\sim 14\%$. The shear rate of the simulation is $500~s^{-1}$.

The npFEM solver has the capability to be executed independently of the fluid solver, using a larger time step, and thus speeding up substantially the overall simulation time. This useful capability is backed by the increased stability of our deformable bodies solver given its implicit nature. Additionally, the implicit Euler introduces a small numerical dissipation which is beneficial for the coupling. The upper limit on the solid solver time step is indirectly imposed by the collision resolution. Large time steps of the npFEM solver may lead to a potential under-resolution of the collisions detection, and thus interpenetrations may be observed, although the overall stability of the collective transport is not affected.

\section{Conclusion}
We presented a novel, highly accurate and robust model for the description of the mechanical, rheological and dynamical properties of red blood cells. The model is characterized by generality, robustness, accuracy and performance, fulfilling all the criteria we consider to be critical aspects for a universal model. To our knowledge, the npFEM solver is a novel approach in the field of computational biophysics to simulate viscoelastic bodies. The solver is independent of mesh resolution or mesh regularity and just one set of parameters can accurately describe the behavior of the body for our full range of case studies. Its extension from a thin-shell model to fully 3D bodies with internal structure and to any type of cells (platelets, white blood cells) is trivial based on the principles discussed at section \ref{sec:npFEM}. The performance of the new continuum-based solver is very close to other state-of-the-art mesoscopic solvers, and thus we managed to bridge the computational gap which was the main obstacle for using the former models.

The framework that combines the npFEM solver, the fluid solver and the FSI is versatile, highly modular and reliable, allowing the user to extend it based on other preferred numerical methods, as well.

As the number of the coupled models is increased, it is not uncommon that the quality of the output decreases. In our case however, extensive validation for many single RBC test cases showed a close match with the experimental data for multiple scenarios.

In the future, based on the proof of capability (see section \ref{sec:ProofOfCapability}), we plan to validate the framework for multiple RBCs test cases and further investigate its scaling with an increasing number of various deformable bodies. Of special interest would be to investigate reduced basis finite elements techniques, that would lead to an even faster deformable bodies solver. Finally, our npFEM solver with its modular interface on evaluating energies and forces could allow an efficient porting of the code to a general-purpose GPU platform.

\section*{Funding}
This project has received funding from the European Union’s Horizon 2020 research and innovation programme under grant agreement No 675451 (CompBioMed project).

\section*{Acknowledgements}
We acknowledge support from the PASC project 2017-2020: Virtual Physiological Blood: an HPC framework for
blood flow simulations in vasculature and in medical devices.

\appendix
\section{Alternating solver} \label{appendix:LocalGlobal}
The minimization of the objective function \eqref{eq:ObjectiveFunction} with strictly quadratic potential energies is performed using an alternating (local/ global) solver. In the local step, $\bm{x}$ is fixed while the projections $\bm{p}$ per element are computed. In the global step, $\bm{p}$ is fixed and the minimum of $g$ is found by equating its gradient to zero:
\begin{equation} \label{eq:gradWithFixedProj}
    \nabla g(\bm{x}_{n+1}) = \frac{1}{h^2} \widetilde{\mathbf{M}} (\bm{x}_{n+1} - \bm{y}_n) + \mathbf{L} \bm{x}_{n+1} - \mathbf{J} \bm{p} = 0,
\end{equation}
so
\begin{equation} \label{eq:x_star}
    \bm{x}_{n+1}^{*} = \left ( \widetilde{\mathbf{M}}/h^2 + \mathbf{L} \right )^{-1} \left ( \mathbf{J} \bm{p} + \widetilde{\mathbf{M}}\bm{y_n}/h^2 \right ).
\end{equation}
The matrix $\widetilde{\mathbf{M}}/h^2 + \mathbf{L}$ is symmetric positive definite and thus $\bm{x}_{n+1}^{*}$ is a global minimum for fixed projections. The above scheme is guaranteed to converge monotonically to a minimum. The local step decreases each constraint keeping fixed the vector $\bm{x}$. The global step, keeping $\bm{p}$ fixed, finds a compromise of the new state towards the projected positions and thus minimizes the objective function. Consequently, this sequence is non-increasing and bounded from below (minimization of Frobenius norms). The local/ global steps are repeated until convergence is satisfied (typically below 10 iterations).

This alternating solver is a \emph{special case of the quasi-Newton method} presented in section \ref{sec:npFEM}. Using the same Hessian approximation $\widetilde{\mathbf{H}} = \widetilde{\mathbf{M}}/h^2 + \mathbf{L}$, it is revealed that the descent direction $\bm{d}$:
\begin{equation}
  \bm{d} = - \left ( \widetilde{\mathbf{M}}/h^2 + \mathbf{L} \right )^{-1} \nabla g(\bm{x}) = \bm{x}^* - \bm{x}.
\end{equation}

Thus, projective dynamics can be viewed as a quasi-Newton method that computes the next iterate as $\bm{x} + \bm{d}$. For normal quasi-Newton methods, we have to use line search techniques to find $\alpha$ such that $\bm{x} + \alpha \bm{d}$ reduces the objective function as much as possible \citep{Nocedal2006NumericalOptimization}. The particularity of PD is that $\alpha = 1$ \citep{Liu2017Quasi-NewtonMaterials} and thus there is no need for line search techniques. At this point, we must highlight that the choice of $\widetilde{\mathbf{H}}$ is inspired by the local/ global solver, as seen from equation \eqref{eq:x_star}, and thus it is deeply rooted in projective dynamics.

\section{FEM reminder} \label{appendix:FEMReminder}
When an object deforms, every material point $\bm{X}$ (undeformed state) is displaced to a new deformed location $\bm{x}$. The deformation function $\bm{\phi} : \mathbb{R}^3 \rightarrow \mathbb{R}^3$ maps every material point to its respective deformed location, such that $\bm{x} = \bm{\phi}(\bm{X})$. The Jacobian matrix of the deformation function $\partial \bm{\phi} / \partial \bm{X}$ is an important physical quantity for describing elastic bodies and is called deformation gradient $\mathbf{F}$. Let us focus on a triangular element isometrically embedded in $\mathbb{R}^2$. The deformation function can be defined to be a \emph{piecewise linear} function $\hat{\bm{\phi}}$ (approximation of $\bm{\phi}$) over the element. In more details, for the triangular elements we have \citep{Sifakis:2012:FSD:2343483.2343501}
\begin{equation} \label{eq:reconstructedDefoMap}
    \hat{\phi}(\bm{X}) = \mathbf{A} \bm{X} + \bm{b},
\end{equation}
where $\mathbf{A}$ \& $\bm{b}$ are element specifics. Differentiating equation \eqref{eq:reconstructedDefoMap} with respect to $\bm{X}$, we conclude that the deformation gradient $\mathbf{F} = \partial \hat{\phi} / \partial \bm{X} = \mathbf{A}$ is constant for the element. Let us denote with $\vec{X}_1,\vec{X}_2,\vec{X}_3$ the undeformed vertex positions of an arbitrary triangular element, and let $\vec{x}_1,\vec{x}_2,\vec{x}_3$ be the deformed counterparts. Each vertex must satisfy the deformation function
\begin{equation}
\begin{Bmatrix}
\vec{x}_1 = \mathbf{F} \vec{X}_1 + \vec{b}\\ 
\vec{x}_2 = \mathbf{F} \vec{X}_2 + \vec{b}\\ 
\vec{x}_3 = \mathbf{F} \vec{X}_3 + \vec{b} 
\end{Bmatrix} \Rightarrow \begin{Bmatrix}
\vec{x}_1 - \vec{x}_3 = \mathbf{F} (\vec{X}_1 - \vec{X}_3) \\ 
\vec{x}_2 - \vec{x}_3 = \mathbf{F} (\vec{X}_2 - \vec{X}_3)
\end{Bmatrix},
\end{equation}
a more compact form reveals that
\begin{align}
\begin{bmatrix}
\vec{x}_1 - \vec{x}_3 & \vec{x}_2 - \vec{x}_3
\end{bmatrix} &= \mathbf{F} \begin{bmatrix}
\vec{X}_1 - \vec{X}_3 & \vec{X}_2 - \vec{X}_3
\end{bmatrix}, \\
\mathbf{D_s} &= \mathbf{F}~\mathbf{D_m}.
\end{align}
Consequently, the deformation gradient $\mathbf{F} \in \mathbb{R}^{2 \times 2}$ of an arbitrary triangular element embedded in $\mathbb{R}^2$ is
\begin{equation}
    \mathbf{F} = \mathbf{D}_s~\mathbf{D}_m^{-1}.
\end{equation}

Of special interest is the derivation of various strain tensors from the deformation gradient. The \emph{Green strain tensor} is given by
\begin{equation}
    \mathbf{E} = \frac{1}{2} \left ( \mathbf{F}^T \mathbf{F} - \mathbf{I} \right ),
\end{equation}
while for small deformations, the \emph{infinitesimal strain tensor} is given by
\begin{equation}
    \bm{\epsilon} = \frac{1}{2} \left ( \mathbf{F} + \mathbf{F}^T \right ) - \mathbf{I}.
\end{equation}

\bibliographystyle{model1-num-names}
\bibliography{Mendeley}

\end{document}